\documentclass[aps,prb]{revtex4}
\usepackage{graphics,epsfig}
\usepackage{amssymb,revsymb}
\usepackage{bm}

\begin{document}
\title{Morphological instability, evolution, and scaling in strained
  epitaxial films: An amplitude equation analysis of the phase field
  crystal model}
\author{Zhi-Feng Huang}
\affiliation{Department of Physics and Astronomy, Wayne State
  University, Detroit, MI 48201}
\author{K. R. Elder}
\affiliation{Department of Physics, Oakland University, Rochester, 
  MI 48309}

\date{\today; to be published in Phys. Rev. B}

\begin{abstract}

Morphological properties of strained epitaxial films are examined
through a mesoscopic approach developed to incorporate both the film
crystalline structure and standard continuum theory. Film surface
profiles and properties, such as surface energy,
liquid-solid miscibility gap and interface thickness, are determined 
as a function of misfit strains and film elastic modulus. We analyze the
stress-driven instability of film surface morphology that leads to the
formation of strained islands.  We find a universal scaling relationship 
between the island size and misfit strain which shows a crossover from the 
well-known continuum elasticity result at the weak strain to a 
behavior governed by a ``perfect" lattice relaxation condition.  
The strain at which the crossover occurs is shown to be a function 
of liquid-solid interfacial thickness, and an asymmetry between 
tensile and compressive strains is observed.
The film instability is found to be accompanied by
mode coupling of the complex amplitudes of
the surface morphological profile, a factor associated with the
crystalline nature of the strained film but absent in conventional
continuum theory.

\end{abstract}


\maketitle

\section{Introduction}

The most recent area of focus in thin film epitaxy has been on exploiting the 
growth and control of strained solid films to develop specific nanostructure
features that can be used in optoelectronic device applications.   
These structures include junctions, quantum wells, and
multilayers/superlattices for which planar interfaces are highly
desired. On the other hand, epitaxially grown films are usually
strained due to the lattice mismatch with the substrate, leading to a
variety of stress-induced effects and structures either on the film
surface or across the interfaces, such as islands (quantum dots) or
nanowires. \cite{re:stangl04,re:shchukin99,re:teichert02,re:berbezier09}
A wide range of device applications results from such heterostructures,
including LEDs, diode lasers, detectors, FETs, etc.,
\cite{re:humphreys08,re:stangl04} with the major technical concerns
being the requirement of long-range ordering, size regularity,
placement and defect control.

Much progress has been made in understanding film growth
above the surface roughening temperature, particularly the formation
and evolution of coherent nanostructures.  The 
evolution sequence often involves many physical processes, including
an initial morphological
instability of the Asaro-Tiller-Grinfeld (ATG) type
\cite{re:asaro72,re:grinfeld86,re:srolovitz89,re:spencer91,re:spencer93}
that results in surface ripples and undulations, 
\cite{re:sutter00,re:tromp00} the formation of islands and the
evolution from pre-pyramid to faceted shape (e.g., $\{105\}$-faceted
pyramids for SiGe \cite{re:tersoff02}), subsequent islands coarsening,
\cite{re:ross98,re:floro00,re:rastelli05} further shape transitions
from pyramids to domes \cite{re:ross98} or to unfaceted prepyramids
\cite{re:rastelli05} and the nucleation of misfit dislocations
for very large islands. \cite{re:jesson95,re:albrecht95}

To understand these complex processes of nanostructure self-assembly,
most of current theoretical efforts are based on either
continuum diffusion and elasticity theories or atomistic simulation
methods that focus on a certain single scale of description. In
standard continuum theory, the film morphology is described by a
coarse-grained, continuum surface profile
\cite{re:srolovitz89,re:spencer91} or phase fields,
\cite{re:muller99,re:kassner01,re:wise05} with evolution governed by
the relaxation of continuum elastic and surface free energies.
Quantitative results have been obtained to reveal fundamental
mechanisms of film nanostructure formation observed in a variety of
experimental systems. Recent work has focused on morphological
instabilities of strained films \cite{re:srolovitz89,re:spencer91,re:spencer93}
or superlattices, \cite{re:shilkrot00,re:huang03a,re:huang03b}
the coupling to alloy film composition inhomogeneity,
\cite{re:guyer95,re:leonard98,re:spencer01,re:huang02a,re:huang02b}
island evolution, \cite{re:spencer05,re:tu07} ordering and coarsening
\cite{re:muller99,re:kassner01,re:wise05,re:liu01,re:levine07,re:huangz07}
as well as island growth on nanomembranes/nanoribbons.
\cite{re:huangm09,re:kimlee09}
Such continuum approaches give a long-wavelength description of 
the system, which has a large computational advantage over microscopic 
approaches but naturally neglects many microscopic crystalline details
that can have a significant impact on film structural evolution and defect
dynamics. This can be remedied via atomistic simulations such as
kinetic Monte Carlo (MC) methods. Recent progress includes identifying 
detailed properties of strained islands such as morphology, density
and size distribution \cite{re:nandipati06,re:zhu07} and 
the evolution of complex surface structures including dots, pits
and grooves as a function of growth conditions in both two 
\cite{re:lam02} and three \cite{re:lung05} dimensions.
However to simulate strained film growth, novel approaches
(e.g., Green's function method \cite{re:lam02,re:zhu07} or
local approximation technique \cite{re:schulze09})
are required to incorporate strain energy via long-range elastic
interactions, which usually limit atomistic
studies to small length and time scales.

Recently an approach coined Phase Field Crystal (PFC) 
modeling has been developed to incorporate atomic-level 
crystalline structures into standard continuum theory for 
pure and binary systems. \cite{re:elder02,re:elder04,re:elder04b,re:elder07}
This model can be related to other continuum field theories such as 
classical density functional 
theory \cite{re:ry79,re:singh91,re:vanteeffelen09,re:kahl09} 
and the atomic density function theory. \cite{re:jin06}  
The PFC model describes the diffusive, large-time-scale dynamics of
the atomic number density field $\rho$, which is spatially periodic on
atomic length scales.  By including atomic scale variations, 
the physics associated with
elasticity, plasticity, multiple crystal orientations and anisotropic
properties (of, e.g., surface energy and elastic constants) is 
naturally incorporated. This approach has been applied to a wide
variety of phenomena including 
glass formation, \cite{re:berry08}
climb and glide dynamics of dislocations, \cite{bme06}
epitaxial growth, \cite{re:elder02,re:elder04,re:elder07,re:huang08,re:wu09,ybv09} 
pre-melting at grain boundaries, \cite{beg08a,mkp08}
commensurate/incommensurate transitions, \cite{ak06,ramos08}
sliding friction phenomena \cite{ak09} and the yield 
strength of polycrystals. \cite{re:elder02,re:elder04,htt09,Ste09} 
For strained film epitaxy, the basic sequence
of film evolution observed in experiments, i.e., morphological
instability $\rightarrow$ nanostructure/island formation $\rightarrow$
dislocation nucleation and climb, has been successfully reproduced in
PFC simulations. \cite{re:elder04,re:elder07,re:huang08,re:wu09}
Unfortunately computational simulations of the original PFC model are 
limited by the need to resolve atomic length scales.  This limitation 
can be overcome by deriving the corresponding amplitude equation
formalism as developed by Goldenfeld 
\textit{et al.} \cite{re:goldenfeld05,re:athreya06} to effectively 
describe the system via the ``slow"-scale amplitude and phase of the
atomic density $\rho$, while at the same time retaining the key 
characters (e.g., elasticity, plasticity and multiple crystal 
orientations) of the modeling.
Very recently such a mesoscopic approach has been extended by Yeon
\textit{et al.} \cite{re:yeon10} to incorporate a slowly-varying
average density field which is essential to account for the
liquid-solid coexistence and a miscibility gap, and also by Elder
\textit{et al.} \cite{re:elder10} to describe the binary alloy systems
for both two-dimensional (2D) hexagonal and three-dimensional (3D) bcc
and fcc structures. Application of this extended expansion to
strained film growth and island formation has yielded promising
results, particularly the determination of a universal size scaling
of surface nanostructures (strained islands). \cite{re:huang08}
However, in these PFC studies some key factors for understanding
the basic mechanisms of strained film evolution are still
missing and yet to be addressed, including film surface properties
(such as strain-dependent surface tension and width) and the effect of 
the sign of film/substrate misfit strain, as will be clarified 
in this work. 

In this paper we provide a complete formulation for such
multiple-scale analysis of single-component, strained film
epitaxy. Compared to our previous work \cite{re:huang08} which is also
based on the amplitude equation formalism established
for two-dimensional high temperature growth, here we provide a
new and more systematic study of various strained film properties
including surface energy, film surface (or liquid-film interface)
thickness, and liquid-film miscibility gap that are identified for
different misfit strains (both tensile and compressive). Furthermore,
morphological instabilities of the strained films and the corresponding
behavior of island formation are systematically investigated, showing
the important effects of misfit strains (both magnitude and sign) and
film surface properties that are absent in previous work. A main
feature of our multi-scale (mesoscopic/microscopic) approach is that
it can maintain the efficiency advantage of the continuum theory
through coarse-grained amplitudes, without losing significant effects
due to the discrete nature of the crystalline film structure.

\section{Amplitude Equation Formalism for Strained Film Epitaxy}

In the PFC model, \cite{re:elder02,re:elder04,re:elder07} the free 
energy functional $F$ can be derived from the classical
density functional theory of freezing \cite{re:elder07} and be
expressed in terms of a dimensionless atomic number density
$n=(\rho-\bar{\rho})/\bar{\rho}$, i.e.,
\begin{equation}
F/\bar{\rho} k_B T = \int d{\bm r}\left \{
\frac{n}{2} \left [ B^\ell+B^x \left ( 2 R^2\nabla^2+R^4\nabla^4
\right) \right ] n -\frac{\tau}{3}n^3 + \frac{v}{4}n^4\right \},
\label{eq:pfc_F}
\end{equation}
where $\bar{\rho}$ is the average density, $T$ is the temperature,
$R$ represents the lattice spacing, $B^\ell$ is related to the
isothermal compressibility of the liquid phase, $B^x$ is proportional
to the bulk modulus of the crystalline state, and $\tau$ and $v$ are
phenomenological parameters (chosen as $\tau=1/2$, $v=1/3$ in the 
following calculations for simplicity). The liquid-solid transition
is controlled by a parameter $\epsilon=(B^x-B^\ell)/B^x$ which is
related to temperature difference from the melting point. The solid
phase exists at $\epsilon>0$, with hexagonal/triangular crystalline
symmetry in 2D and bcc in 3D. Based on the assumption of conserved 
system dynamics, i.e., $\partial n / \partial t = \Gamma \nabla^2
\delta F / \delta n$ with $\Gamma$ the mobility, the PFC
dynamic equation is given by
\begin{equation}
\partial n / \partial t = \Gamma \nabla^2 \left [ B^\ell n + B^x (R^4
  \nabla^4 + 2R^2 \nabla^2) n - \tau n^2 + v n^3 \right ].
\label{eq:pfc0}
\end{equation}
Defining a length scale $l_0=R$, a time scale
$\tau_0=R^2/\Gamma B^x$, and $n \rightarrow \sqrt{v/B^x}~n$, we
obtain the rescaled equation
\begin{equation}
\partial n / \partial t = \nabla^2 \left [ -\epsilon\,n + (\nabla^2 +
  q_0^2)^2 n - g n^2 +n^3 \right ],
\label{eq:pfc}
\end{equation}
where $g=\tau / \sqrt{v B^x}$, $q_0=1$ and the symbol
$q_0$ is retained for the clarity of presentation.

For the epitaxial system of interest, we consider a system
configuration composed of a semi-infinite strained crystalline film
and a coexisting homogeneous liquid state, which are 
separated by a time-evolving interface (i.e., film surface). 
To access the ``slow" time and length scales of the film surface
profile we introduce a standard multiple scale expansion 
of the PFC equation (\ref{eq:pfc}) and derive the associated 
amplitude equations, with detailed procedures given in Refs. 
\onlinecite{re:goldenfeld05,re:athreya06,re:yeon10}.
For a 2D system with the film surface normal to the $y$ direction,
the atomic density field $n$ is expanded in both liquid and solid
regions as the superposition of a spatially/temporally-varying
average local density $n_0$ (for the zero wavenumber mode) and three
hexagonal base modes, i.e.,
\begin{equation}
n = n_0(X,Y,T)+\sum\limits_{j=1}^{3} A_j(X,Y,T) e^{i{\bm
    q}_j^0 \cdot {\bm r}} + {\rm c.c.},
\label{eq:nexpan}
\end{equation}
where both $n_0$ and complex amplitudes $A_j$ are slowly varying 
variables (with $A_j=0$ in the liquid region), 
and ${\bm q}_j^0$ represent the three hexagonal basic wave vectors 
\begin{equation}
{\bm q_1^0} = q_0 \left ( -\frac{\sqrt{3}}{2} \hat{x} - \frac{1}{2}
\hat{y} \right ), \quad {\bm q_2^0} = q_0 \hat{y}, \quad {\bm q_3^0} =
q_0 \left ( \frac{\sqrt{3}}{2} \hat{x} - \frac{1}{2} 
\hat{y} \right ).
\label{eq:qj0}
\end{equation}
This expansion (\ref{eq:nexpan}) implies the separation of
``slow" scales
$X=\epsilon^{1/2} x$, $Y=\epsilon^{1/2} y$, $T=\epsilon t$
for $n_0$ and $A_j$ (and hence the film surface profile) from the
underlying crystalline structure, at the limit of small $\epsilon$
or high temperature growth. The corresponding amplitude equations
are given by (in the form of Model C \cite{re:hohenberg77})
\begin{eqnarray}
&\partial A_j / \partial t =  - q_0^2 \delta {\cal F} / \delta A_j^*,& 
\label{eq:amplA}\\
&\partial n_0 / \partial t = \nabla^2 \delta {\cal F} / \delta n_0,&
\label{eq:amplB}
\end{eqnarray}
where the effective potential ${\cal F}$ (a Lyapunov functional) is 
written as
\begin{eqnarray}
&{\cal F} =& \int d {\bm r} \left \{ (-\epsilon +3n_0^2-2gn_0) 
\sum_{j=1}^3 |A_j|^2 + \sum_{j=1}^3 \left | \left ( \nabla^2 
+ 2i{\bm q}_j^0 \cdot {\bm \nabla} \right ) A_j \right |^2  
+ \frac{3}{2}\sum_{j=1}^3 |A_j|^4 \right. \nonumber\\
&& + (6n_0-2g) (A_1A_2A_3+A_1^*A_2^*A_3^*)
+ 6\left ( |A_1|^2 |A_2|^2 + |A_1|^2 |A_3|^2 + |A_2|^2 |A_3|^2
\right ) \nonumber\\
&& \left. -\frac{1}{2} \epsilon n_0^2 + \frac{1}{2} \left [ \left (
  \nabla^2 + q_0^2 \right ) n_0 \right ]^2 - \frac{1}{3} g n_0^3 +
\frac{1}{4} n_0^4 \right \}. \label{eq:F}
\end{eqnarray}
Note that the operator 
$(\nabla^2 + 2i{\bm q}_j^0 \cdot {\bm \nabla})$ preserves the 
rotational covariance of these amplitude
equations. \cite{re:gunaratne94}  This effective free energy describes
a first order phase transition from a liquid ($A_j=0$) to 
a solid state ($A_j\neq 0$) and incorporates elasticity 
though the operator $(\nabla^2 + 2i{\bm q}_j^0 \cdot {\bm \nabla})$,
as discussed in Ref. \onlinecite{re:elder10}.  In addition the terms
containing $n_0$ incorporate a miscibility gap for the density at
liquid-solid coexistence. 

For a hexagonal lattice, the equilibrium wave numbers along $x$ and $y$
directions are $q_{x_0}=\sqrt{3} q_0/2$ and $q_{y_0}=q_0$ for the
undistorted, zero-misfit bulk lattice. For strained films during
epitaxy (with distorted hexagons/triangles), the misfit $\varepsilon_m$
is determined by
\begin{equation}
\varepsilon_m = \frac{a_0-a}{a} = \frac{q_x}{q_{x_0}}-1,
\label{eq:misfit}
\end{equation}
where $a_0=2\pi/q_{x_0}$ is the stress-free bulk film lattice constant
and $a=2\pi/q_x$ is the lattice constant of the strained film. The
complex amplitudes $A_j$ should then be expressed by
\begin{equation}
A_1 = A_1' e^{-i(\delta_x x + \delta_y y /2)}, \quad A_2 = A_2' e^{i
  \delta_y y}, \quad A_3 = A_3' e^{i(\delta_x x - \delta_y y /2)},
\label{eq:A'}
\end{equation}
where amplitudes $A_j'$ are complex,
$\delta_x = q_{x_0} \varepsilon_m = \sqrt{3} q_0 \varepsilon_m /2$,
and the value of $\delta_y$ ($\neq \delta_x$) is determined by the lattice
relaxation along the film growth direction $y$ (corresponding to the 
Poisson relaxation in continuum elasticity theory).
Since both $A_j$ and $A_j'$ are slowly varying quantities,
$\delta_x$, $\delta_y$ and the misfit strain ($\varepsilon_m$) should also be
sufficiently small. Substituting Eq. (\ref{eq:A'}) into 
Eqs. (\ref{eq:amplA})--(\ref{eq:F}), 
the amplitude equations for strained films are then
\begin{eqnarray}
&\partial_t A_1' =& -q_0^2 \left \{ \left [ -\epsilon + 3n_0^2-2gn_0 +
  \left ( \partial_x^2+\partial_y^2 - i (\sqrt{3} q_0 +
  2\delta_x)\partial_x -i (q_0 + \delta_y) \partial_y
  \right. \right. \right. \nonumber\\
  && \left. \left. -\sqrt{3} q_0
  \delta_x - \delta_x^2 - q_0 \delta_y /2 -\delta_y^2 /4 \right )^2
  \right ] A_1' +(6n_0-2g) A_2'^* A_3'^* \nonumber\\
&&  \left. + 3A_1' \left ( |A_1'|^2 +
      2|A_2'|^2 + 2|A_3'|^2 \right ) \right \} ,
\label{eq:amplA1'}\\
&\partial_t A_2' =& -q_0^2 \left \{ \left [ -\epsilon + 3n_0^2-2gn_0 +
  \left ( \partial_x^2+\partial_y^2 + 2i (q_0 + \delta_y) \partial_y
  - 2q_0 \delta_y -\delta_y^2 \right )^2
  \right ] A_2' \right. \nonumber\\
&&  \left. +(6n_0-2g) A_1'^* A_3'^* + 3A_2' \left ( |A_2'|^2 +
      2|A_1'|^2 + 2|A_3'|^2 \right ) \right \} ,
\label{eq:amplA2'}\\
&\partial_t A_3' =& -q_0^2 \left \{ \left [ -\epsilon + 3n_0^2-2gn_0 +
  \left ( \partial_x^2+\partial_y^2 + i (\sqrt{3} q_0 +
  2\delta_x)\partial_x -i (q_0 + \delta_y) \partial_y
  \right. \right. \right. \nonumber\\
  && \left. \left. -\sqrt{3} q_0
  \delta_x - \delta_x^2 - q_0 \delta_y /2 -\delta_y^2 /4 \right )^2
  \right ] A_3' +(6n_0-2g) A_1'^* A_2'^* \nonumber\\
&&  \left. + 3A_3' \left ( |A_3'|^2 +
      2|A_1'|^2 + 2|A_2'|^2 \right ) \right \} ,
\label{eq:amplA3'}\\
&\partial_t n_0  =& \nabla^2 \left \{ \left [ -\epsilon + \left ( 
      \nabla^2 + q_0^2 \right )^2 \right ] n_0 -g n_0^2  + n_0^3  
      + (6n_0-2g) \left ( |A_1'|^2 + |A_2'|^2 + |A_3'|^2 \right ) \right .
\nonumber\\
&& \left. +6(A_1'A_2'A_3'+A_1'^*A_2'^*A_3'^*) \right \}. \label{eq:ampln0'}
\end{eqnarray}

	These amplitude equations describe a strained system and 
will be used to study morphological instabilities of a liquid-crystal 
surface under strain.  In the next section, steady state or base solutions 
will be obtained for a planar liquid-crystal interface under strain.  
In Sec. \ref{sec:perb} the stability of these planar solutions to 
small perturbations at the surface  will be examined.

\section{Base State Solution: Film Surface Properties}
\label{sec:base}

We first construct a base state involving a planar film surface (i.e.,
a coexisting liquid-crystal interface). The corresponding
amplitudes $A_j^0$ and density $n_0^0$ are then only a function of the 
normal direction $y$, and hence the amplitude equations
(\ref{eq:amplA1'})--(\ref{eq:ampln0'}) can be simplified as
\begin{equation}
\partial A_j^0 / \partial t =  - q_0^2 \delta {\cal F}^0 / \delta
  {A_j^0}^*, \qquad
\partial n_0^0 / \partial t = \partial_y^2 \delta {\cal F}^0 / \delta
  n_0^0, \label{eq:An0}
\end{equation}
where
\begin{eqnarray}
&{\cal F}^0 =& \int d {\bm r} \left \{ (-\epsilon +3{n_0^0}^2-2gn_0^0)
\sum_{j=1}^3 |A_j^0|^2 + \frac{3}{2}\sum_{j=1}^3 |A_j^0|^4 \right. \nonumber\\
&& + \left |  \left [ \partial_y^2 -i (q_0 + \delta_y) \partial_y
-\sqrt{3} q_0 \delta_x - \delta_x^2 - q_0 \delta_y /2 -\delta_y^2 /4
\right ] A_1^0 \right |^2 \nonumber\\
&& + \left |  \left [ \partial_y^2 + 2i (q_0 + \delta_y) \partial_y
- 2q_0 \delta_y -\delta_y^2 \right ] A_2^0 \right |^2 \nonumber\\
&& + \left |  \left [ \partial_y^2 -i (q_0 + \delta_y) \partial_y
-\sqrt{3} q_0 \delta_x - \delta_x^2 - q_0 \delta_y /2 -\delta_y^2 /4
\right ] A_3^0 \right |^2 \nonumber\\
&& + (6n_0^0-2g) (A_1^0A_2^0A_3^0+{A_1^0}^*{A_2^0}^*{A_3^0}^*)
+ 6\left ( |A_1^0|^2 |A_2^0|^2 + |A_1^0|^2 |A_3^0|^2 + |A_2^0|^2 |A_3^0|^2
\right ) \nonumber\\
&& \left. -\frac{1}{2} \epsilon {n_0^0}^2 + \frac{1}{2} \left [ \left (
  \partial_y^2 + q_0^2 \right ) n_0^0 \right ]^2 - \frac{1}{3} g {n_0^0}^3 +
\frac{1}{4} {n_0^0}^4 \right \}. \label{eq:F0}
\end{eqnarray}

\begin{figure}
\centerline{
\includegraphics[height=2.8in]{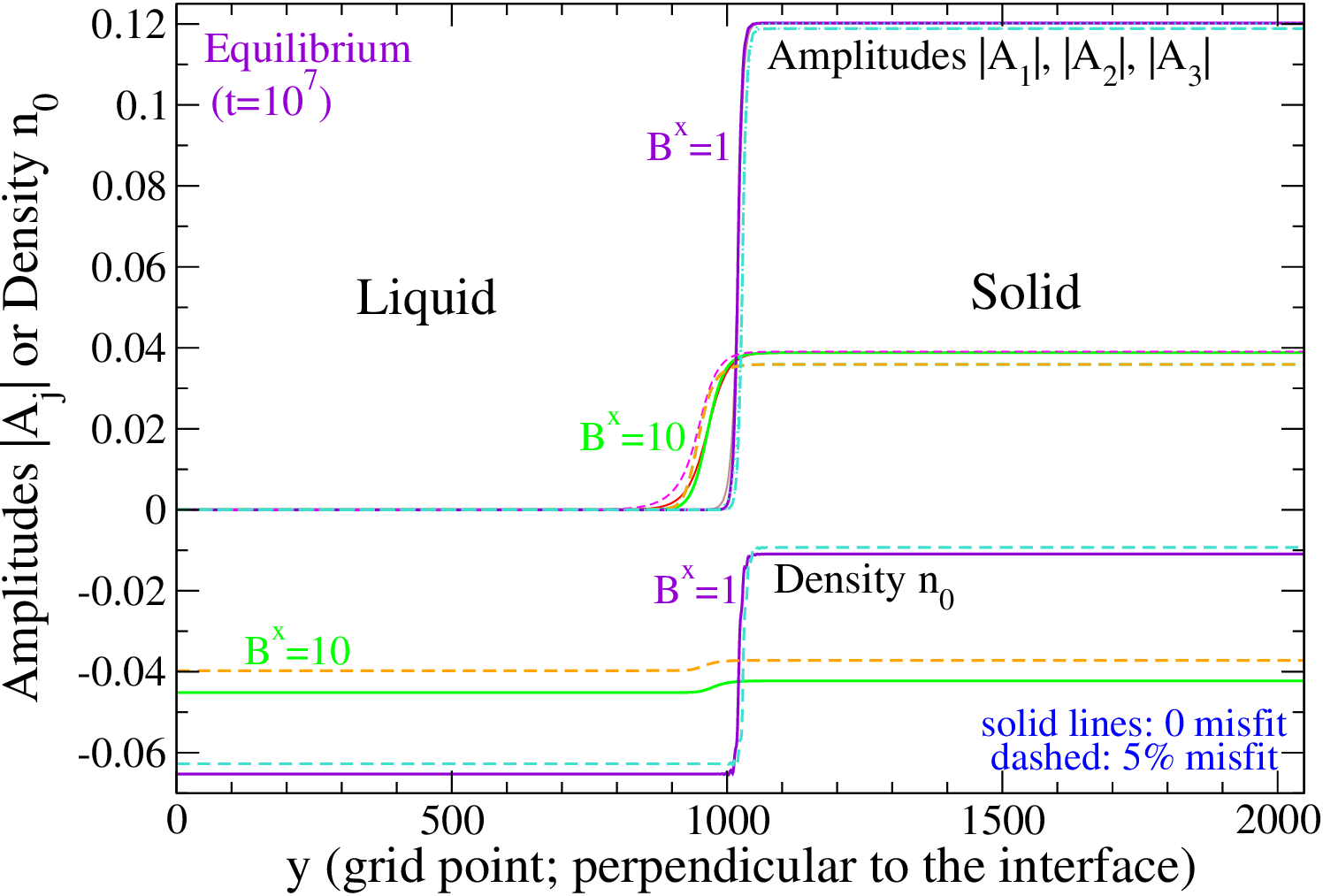}}
\caption{The equilibrium (solid/liquid coexistence) profile of the
  base state, for $\epsilon=0.02$, $B^x=1$ and $10$, and misfit
  $\varepsilon_m =0$ (solid lines) and $5\%$ (dashed lines).}
\vskip 20pt
\label{fig:A0n0}
\end{figure}

\begin{figure}
\centerline{
\includegraphics[height=2.8in]{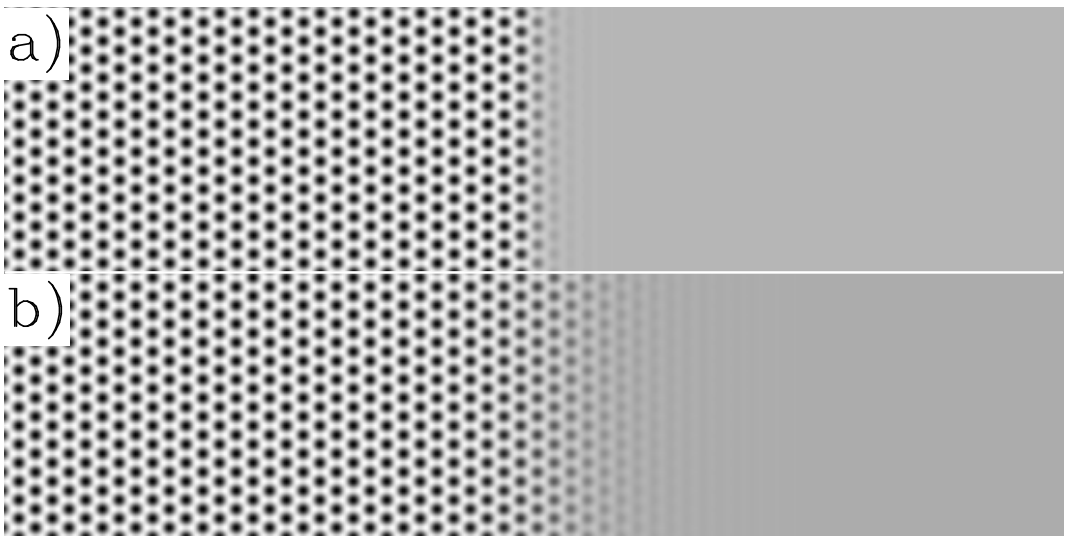}}
\caption{Sample equilibrium profiles of the complete density field $n$
as reconstructed from $n_0^0$ and $A_j^0$ for $\epsilon=0.02$, a misfit 
of $3\%$, and $B^x=1$ and $10$ in a) and b) respectively.}
\label{fig:EqI}
\end{figure}

\begin{figure}
\centerline{
\includegraphics[height=2.8in]{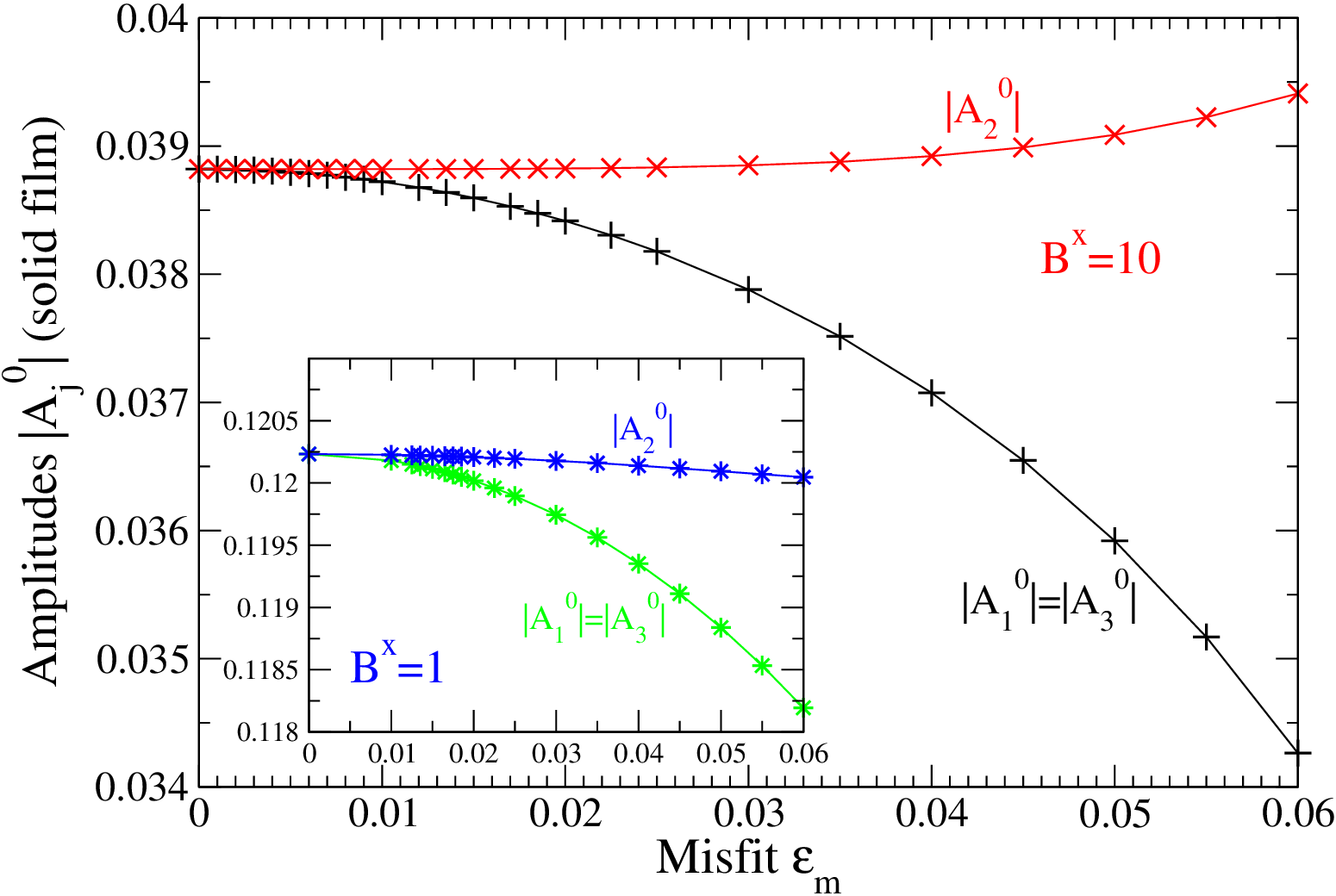}}
\caption{The equilibrium amplitudes $|A_j^0|$ in the solid region as a
  function of misfit $\varepsilon_m$, for $\epsilon=0.02$. The results
  in the main panel correspond to $B^x=10$, and those in the inset are
  for $B^x=1$. Note the much larger vertical scale in the inset.}
\label{fig:A0}
\end{figure}

The equilibrium profile for the base state (with solid/liquid
coexistence) is given in Fig. \ref{fig:A0n0}, corresponding to
non-growing, stationary films of different misfit strains $\varepsilon_m$ 
and elastic constants (as determined by $B^x$).  The amplitudes and 
$n_0^0$ can be used to reconstruct the full density field $n$ via 
Eq. (\ref{eq:nexpan}), as shown in Fig. \ref{fig:EqI}.  
This figure highlights the increase in interfacial width 
as the magnitude of elastic moduli (i.e., $B^x$) increases.
Since the stationary solution of Eqs. (\ref{eq:An0}) and (\ref{eq:F0})
cannot be obtained analytically, the results 
shown were obtained by numerical
solutions based on a pseudospectral method. To apply the periodic
boundary condition, we set the initial configuration as a pair of
symmetric liquid-solid interfaces located at $y=L_y/4$ and $3 L_y/4$
respectively, with $L_y$ the one-dimensional (1D) system size which
is chosen up to $L_y=8192$ in our calculations so that these two
interfaces are sufficiently far apart from each other and thus
evolve independently. In the numerical algorithm adopted, the
second order Crank-Nicholson time stepping scheme is used for the 
linear terms, while a second order Adams-Bashford explicit method
is applied for the nonlinearities. A grid spacing $\Delta y = \lambda_0/8$
(i.e., 8 grid points per basic wavelength $\lambda_0=2\pi /q_0$) is 
chosen in most of calculations, although similar results have been obtained
with much larger $\Delta y$. Relatively large time steps $\Delta t$ can 
be adopted without losing numerical stability: We use $\Delta t = 0.5$
(or even $1$) for $B^x \geq 10$, and $\Delta t = 0.2$ for $B^x=1$ with
sharp interface. We also use the same algorithm and parameters in the 
stability/perturbation calculations given in Sec. \ref{sec:perb}.

\begin{figure}
\vskip 0.3in
\centerline{
\includegraphics[height=2.3in]{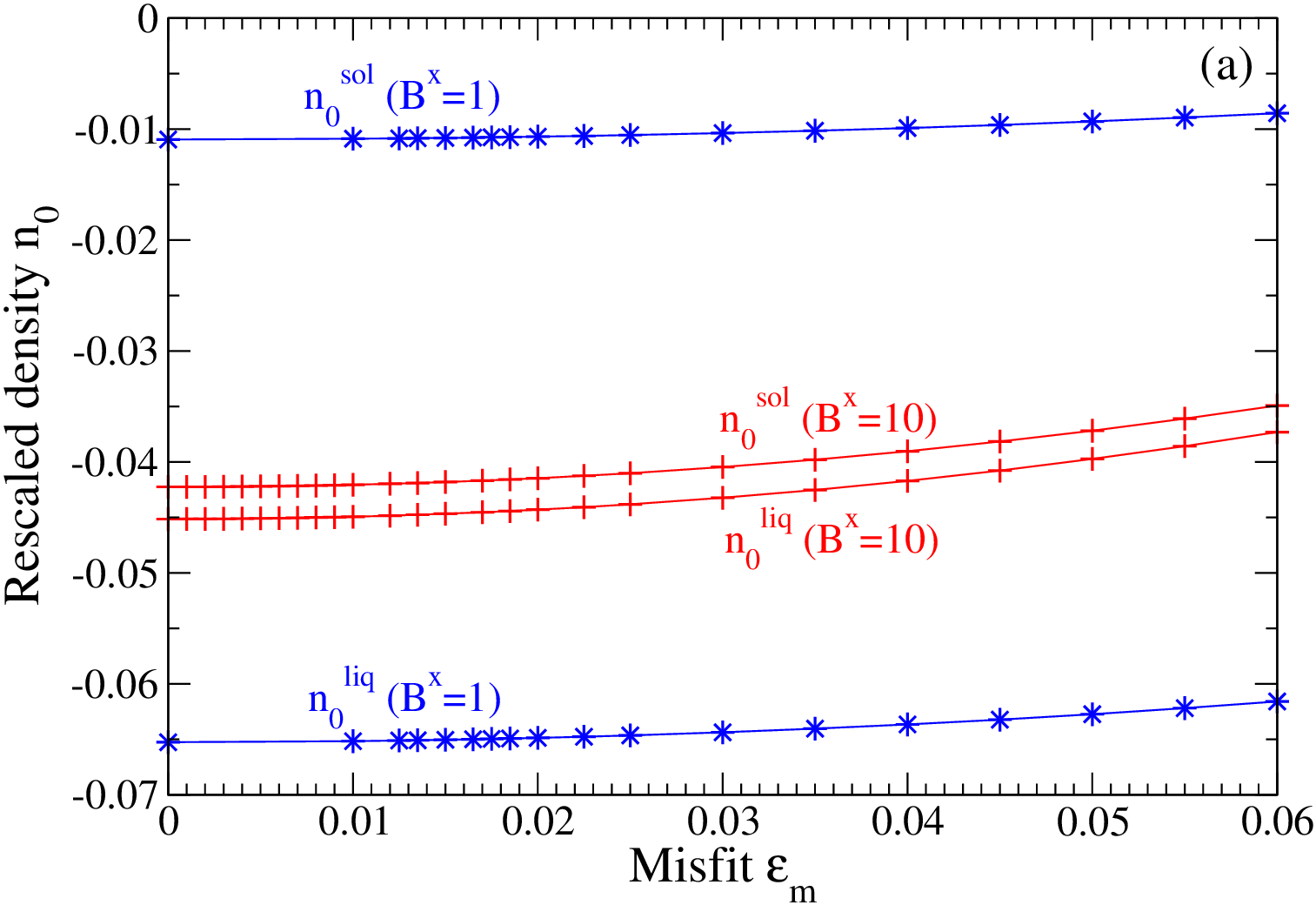} \hfill
\includegraphics[height=2.3in]{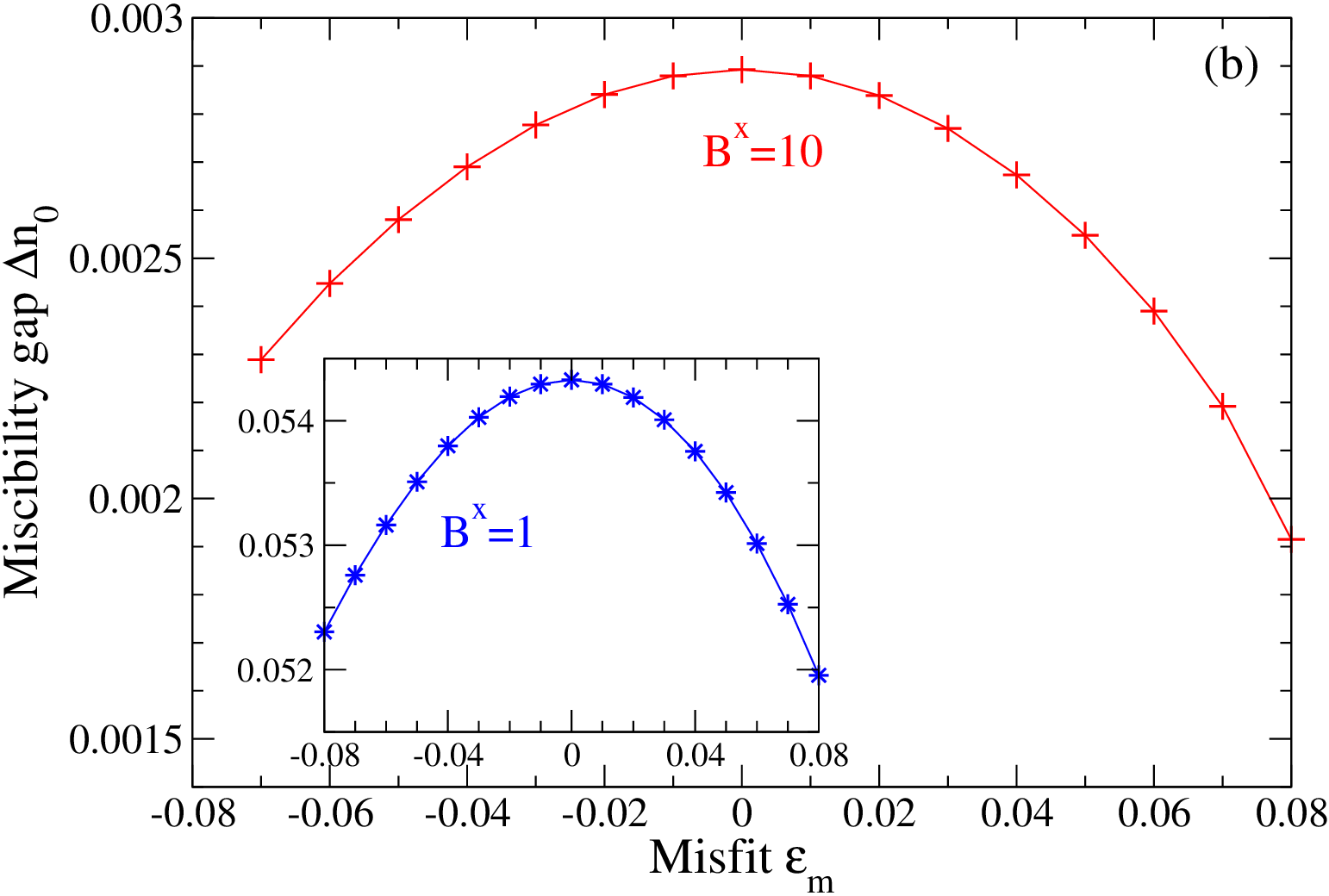}}
\caption{(a) The equilibrium densities $n_0^{\rm liq}$ and $n_0^{\rm
  sol}$ (in the coexisting liquid and solid regions respectively) 
  as a function of misfit strain $\varepsilon_m$, 
  with parameters the same as those of Fig. \ref{fig:A0}.
  (b) The size of miscibility gap $\Delta n_0 = n_0^{\rm sol} -
  n_0^{\rm liq}$ as a function of $\varepsilon_m$, for $B^x=10$ (the
  main panel) and $1$ (the inset); Note the large vertical scale for
  $B^x=1$ in the inset.} 
\label{fig:n0}
\end{figure}

For finite misfits the amplitudes $|A_1^0|=|A_3^0| \neq
|A_2^0|$ and their difference increases with $\varepsilon_m$ 
as shown in Fig. \ref{fig:A0}. This corresponds to a triangular structure 
distorted along the $y$ direction (the surface normal) and the 
degree of distortion increases with misfit strain.
Also as shown in Fig. \ref{fig:A0n0}, for larger value of 
$B^x$ which corresponds to smaller bulk modulus (as we calculate
based on one-mode approximation; see Sec. \ref{sec:perb}), 
the interface or film surface is more diffuse 
(i.e., with larger interface width), but with a narrower 
coexistence region (i.e., smaller but nonzero miscibility gap). 
This can also be seen in Fig. \ref{fig:n0}, which shows the liquidus 
and solidus rescaled density $n_0^{\rm liq}$, $n_0^{\rm sol}$ as
well as the miscibility gap $\Delta n_0 =n_0^{\rm sol} - n_0^{\rm liq}$ 
as a function of misfit $\varepsilon_m$. The size of miscibility gap
decreases with the increasing magnitude of misfit, and shows slight
asymmetry with respect to the misfit sign as a result of different
non-linear elastic effects on liquid-solid coexistence property for
tensile and compressive strains. 

\begin{figure}
\vskip .3in
\centerline{
\includegraphics[height=2.8in]{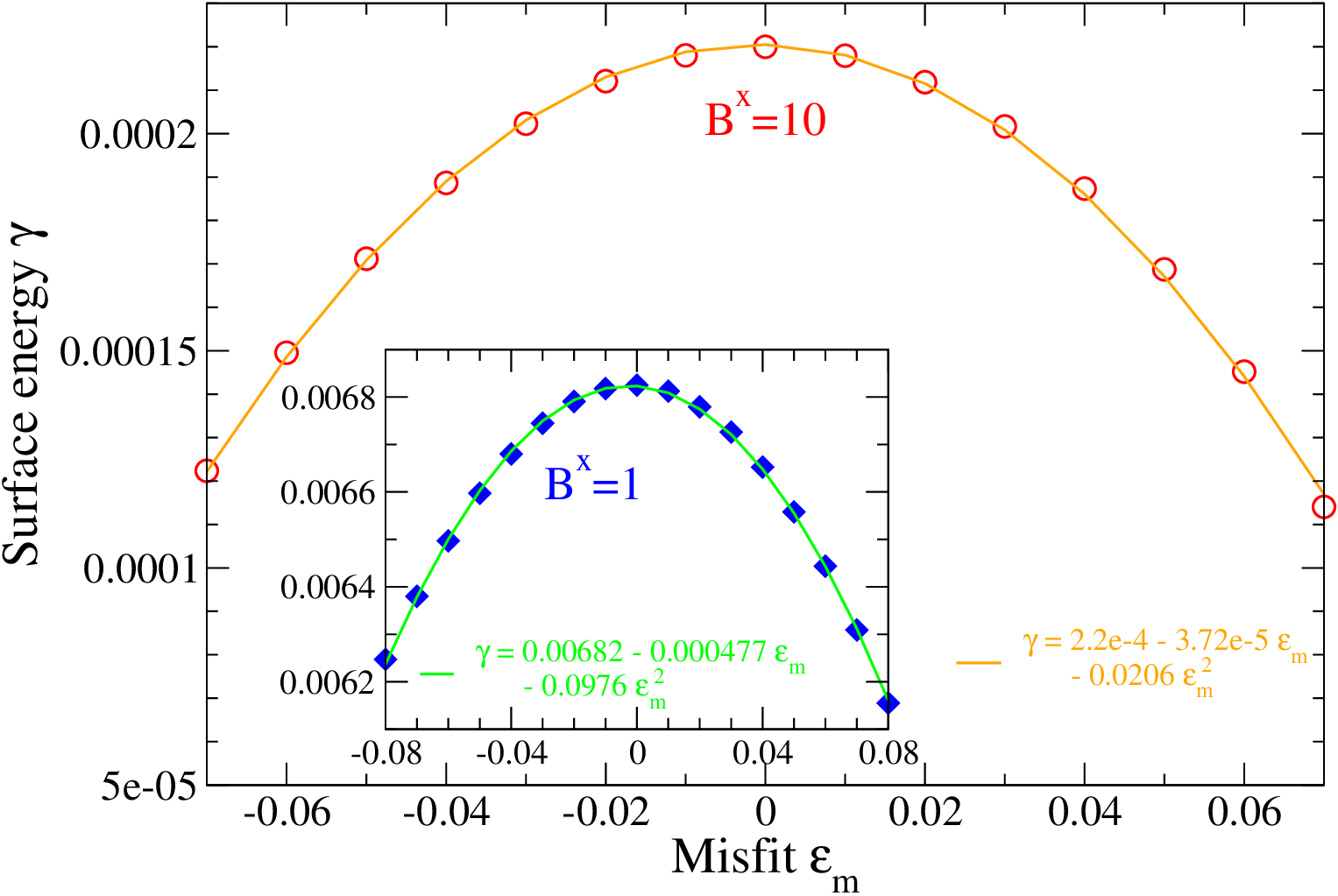}}
\caption{Results of surface tension $\gamma$ for different misfit
strains $\varepsilon_m$, with parameters the same as those of 
Fig. \ref{fig:A0}. Also shown are the quadratic fitting results
for $B^x=10$ (in the main panel) and $1$ (in the inset). Note that the
vertical scale in the inset for $B^x=1$ is much larger.}
\label{fig:gamma}
\end{figure}

We also calculate the surface tension $\gamma$ as a function of
misfit strain since it is one of the important factors for determining 
film stability and island formation. Surface energy is known
to play a stabilization role on film evolution and for simplicity
is often approximated as misfit independent in many strained film studies.
\cite{re:srolovitz89,re:spencer91,re:spencer93,re:muller99,re:kassner01,%
re:wise05,re:shilkrot00,re:huang03a,re:huang03b,re:guyer95,re:leonard98,%
re:spencer01,re:huang02a,re:huang02b}
However in the presence of a strain field, the surface energy is known 
to vary as a result of intrinsic surface stress ${\bm \sigma}^0$ and is
usually expanded up to 2nd order in terms of strain tensor $u_{ij}$
(with $i,j$ the film surface coordinate indices) in linear elasticity
theory, \cite{re:wolf93,re:shchukin99} i.e., 
\begin{equation}
\gamma = \gamma_0 + \sigma^0_{ij} u_{ij} + \frac{1}{2} S_{ijkl} u_{ij} u_{kl},
\label{eq:gamma_expan}
\end{equation}
where $S_{ijkl}$ are the surface excess elastic moduli.  Both 
$\sigma_{ij}^0$ and $S_{ijkl}$ can be either positive or 
negative. \cite{re:shchukin99}
For the 1D surface considered here, strain $u_{xx} = 
\varepsilon_m$ and hence Eq. (\ref{eq:gamma_expan}) gives 
$\gamma = \gamma_0 + \sigma^0_{xx} \varepsilon_m + S_{xxxx} 
\varepsilon_m^2 /2$, which is consistent with our amplitude-equation
calculations shown in Fig. \ref{fig:gamma}. Data fitting of our
numerical results yields $\gamma_0=6.82 \times 10^{-3}$,
$\sigma^0_{xx}=-4.77 \times 10^{-4}$, $S_{xxxx}/2=-9.76 \times 10^{-2}$
for $B^x=1$, and $\gamma_0=2.20 \times 10^{-4}$, $\sigma^0_{xx}
= -3.72 \times 10^{-5}$, $S_{xxxx}/2=-2.06 \times 10^{-2}$
for $B^x=10$ (all in dimensionless unit), showing smaller surface 
energy for larger value of $B^x$ (with larger surface width).
These results indicate that for the parameters chosen, 
both the intrinsic surface stress $\sigma^0_{xx}$ and excess elastic moduli $S_{xxxx}$
are negative, leading to the decrease of surface energy with increasing
magnitude of misfit strain.  In addition the tensile surface stress is rather 
weak which can explain the weak asymmetry of $\gamma$ between tensile 
and compressive strained films.

\section{Morphological Instability and Island Scaling}
\label{sec:perb}

For strained films with nonzero misfit, a morphological instability of film
surface is known to develop as a result of strain energy relaxation,
leading to surface undulations and then the formation of surface
nanostructures such as strained islands. Such an instability can be 
revealed via a linear analysis of amplitude equations given above.
We can expand the amplitudes in Fourier series as
\begin{eqnarray}
& A_j'(x,y,t) = A_j^0(y) + \sum\limits_{q_x} \hat{A}_j(q_x,y,t) e^{i q_x
    x},& \label{eq:A_expan}\\
& n_0(x,y,t) = n_0^0(y) + \sum\limits_{q_x} \hat{n}_0(q_x,y,t) e^{i q_x x},&
\label{eq:n_expan}
\end{eqnarray}
where $A_j^0(y)$ and $n_0^0(y)$ are the planar base solutions discussed 
in the previous section and the perturbed quantities 
$\hat{A}_j$ and $\hat{n}_0$ obey the following linearized equations, 
\begin{eqnarray}
&\partial_t \hat{A}_1(q_x,y,t)& = -q_0^2 \left \{ 
  \left [ -\epsilon + 3{n_0^0}^2-2gn_0^0 +
  \left ( \partial_y^2 -i (q_0 + \delta_y) \partial_y - q_x^2 
  + (\sqrt{3} q_0 + 2\delta_x)q_x
  \right. \right. \right. \nonumber\\
&& \left. \left. -\sqrt{3} q_0
  \delta_x - \delta_x^2 - q_0 \delta_y /2 -\delta_y^2 /4 \right )^2
  + 6 \left ( |A_1^0|^2 + |A_2^0|^2 + |A_3^0|^2 \right ) \right ]
  \hat{A}_1(q_x,y,t) \nonumber\\
&& + 6A_1^0 \left [ {A_2^0}^* \hat{A}_2(q_x,y,t) + {A_3^0}^*
  \hat{A}_3(q_x,y,t) \right ] + 3{A_1^0}^2 \hat{A}_1^*(-q_x,y,t) \nonumber\\
&& + \left [ (6n_0^0-2g){A_3^0}^* + 6A_1^0A_2^0 \right ]
  \hat{A}_2^*(-q_x,y,t) \nonumber\\
&& + \left [ (6n_0^0-2g){A_2^0}^* + 6A_1^0A_3^0 \right ]
  \hat{A}_3^*(-q_x,y,t) \nonumber\\
&&  \left. + \left [ (6n_0^0-2g)A_1^0 + 6{A_2^0}^*{A_3^0}^* \right ]
  \hat{n}_0(q_x,y,t) \right \},
\label{eq:ampl_hatA1}\\
&\partial_t \hat{A}_2(q_x,y,t)& = -q_0^2 \left \{ 
  \left [ -\epsilon + 3{n_0^0}^2-2gn_0^0 +
  \left ( \partial_y^2 + 2i (q_0 + \delta_y) \partial_y - q_x^2 
  \right. \right. \right. \nonumber\\
&& \left. \left. - 2q_0 \delta_y -\delta_y^2 \right )^2
  + 6 \left ( |A_1^0|^2 + |A_2^0|^2 + |A_3^0|^2 \right ) \right ]
  \hat{A}_2(q_x,y,t) \nonumber\\
&& + 6A_2^0 \left [ {A_1^0}^* \hat{A}_1(q_x,y,t) + {A_3^0}^*
  \hat{A}_3(q_x,y,t) \right ] + 3{A_2^0}^2 \hat{A}_2^*(-q_x,y,t) \nonumber\\
&& + \left [ (6n_0^0-2g){A_3^0}^* + 6A_1^0A_2^0 \right ]
  \hat{A}_1^*(-q_x,y,t) \nonumber\\
&& + \left [ (6n_0^0-2g){A_1^0}^* + 6A_2^0A_3^0 \right ]
  \hat{A}_3^*(-q_x,y,t) \nonumber\\
&&  \left. + \left [ (6n_0^0-2g)A_2^0 + 6{A_1^0}^*{A_3^0}^* \right ]
  \hat{n}_0(q_x,y,t) \right \},
\label{eq:ampl_hatA2}\\
&\partial_t \hat{A}_3(q_x,y,t)& = -q_0^2 \left \{ 
  \left [ -\epsilon + 3{n_0^0}^2-2gn_0^0 +
  \left ( \partial_y^2 -i (q_0 + \delta_y) \partial_y - q_x^2 
  - (\sqrt{3} q_0 + 2\delta_x)q_x
  \right. \right. \right. \nonumber\\
&& \left. \left. -\sqrt{3} q_0
  \delta_x - \delta_x^2 - q_0 \delta_y /2 -\delta_y^2 /4 \right )^2
  + 6 \left ( |A_1^0|^2 + |A_2^0|^2 + |A_3^0|^2 \right ) \right ]
  \hat{A}_3(q_x,y,t) \nonumber\\
&& + 6A_3^0 \left [ {A_1^0}^* \hat{A}_1(q_x,y,t) + {A_2^0}^*
  \hat{A}_2(q_x,y,t) \right ] + 3{A_3^0}^2 \hat{A}_3^*(-q_x,y,t) \nonumber\\
&& + \left [ (6n_0^0-2g){A_2^0}^* + 6A_1^0A_3^0 \right ]
  \hat{A}_1^*(-q_x,y,t) \nonumber\\
&& + \left [ (6n_0^0-2g){A_1^0}^* + 6A_2^0A_3^0 \right ]
  \hat{A}_2^*(-q_x,y,t)  \nonumber\\
&&  \left. + \left [ (6n_0^0-2g)A_3^0 + 6{A_1^0}^*{A_2^0}^* \right ]
  \hat{n}_0(q_x,y,t) \right \},
\label{eq:ampl_hatA3}\\
&\partial_t \hat{n}_0(q_x,y,t)& = \left ( \partial_y^2 - q_x^2 \right ) 
  \left \{ \left [ -\epsilon + 3{n_0^0}^2-2gn_0^0 + \left ( \partial_y^2
   -q_x^2 + q_0^2 \right )^2 \right. \right. \nonumber\\ 
&& \left. + 6 \left ( |A_1^0|^2 + |A_2^0|^2 + |A_3^0|^2 \right )
  \right ] \hat{n}_0(q_x,y,t) \nonumber\\
&& + \left [ (6n_0^0-2g){A_1^0}^* + 6A_2^0A_3^0 \right ]
  \hat{A}_1(q_x,y,t) \nonumber\\
&& + \left [ (6n_0^0-2g){A_2^0}^* + 6A_1^0A_3^0
  \right ] \hat{A}_2(q_x,y,t) \nonumber\\
&& + \left [ (6n_0^0-2g){A_3^0}^* + 6A_1^0A_2^0 \right ]
  \hat{A}_3(q_x,y,t) \nonumber\\
&& + \left [ (6n_0^0-2g)A_1^0 + 6{A_2^0}^*{A_3^0}^* \right ]
  \hat{A}_1^*(-q_x,y,t) \nonumber\\
&& + \left [ (6n_0^0-2g)A_2^0 + 6{A_1^0}^*{A_3^0}^* \right ]
  \hat{A}_2^*(-q_x,y,t) \nonumber\\ 
&& \left. + \left [ (6n_0^0-2g)A_3^0 + 6{A_1^0}^*{A_2^0}^* \right ]
  \hat{A}_3^*(-q_x,y,t)  \right \}. \label{eq:ampl_hatn0}
\end{eqnarray}

The stability of the base planar film surface is examined by
introducing initial small random perturbations into $\hat{A}_j$ and
$\hat{n}_0$, and solving numerically the initial value problem defined 
by Eqs. (\ref{eq:ampl_hatA1})--(\ref{eq:ampl_hatn0}), given a specific 
value of $q_x$. The numerical algorithm introduced in 
Sec. \ref{sec:base} is employed, with the use of a pseudospectral method and 
periodic boundary conditions. 

\begin{figure}
\centerline{
\includegraphics[height=1.8in]{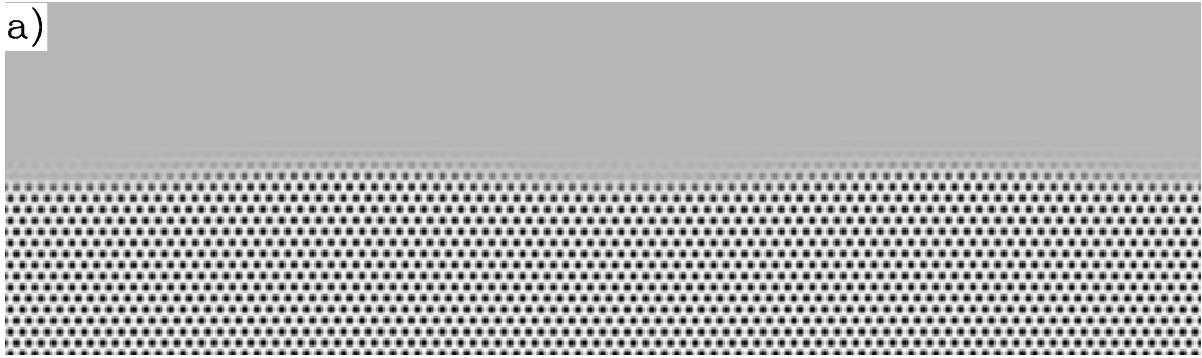}}
\centerline{
\includegraphics[height=1.8in]{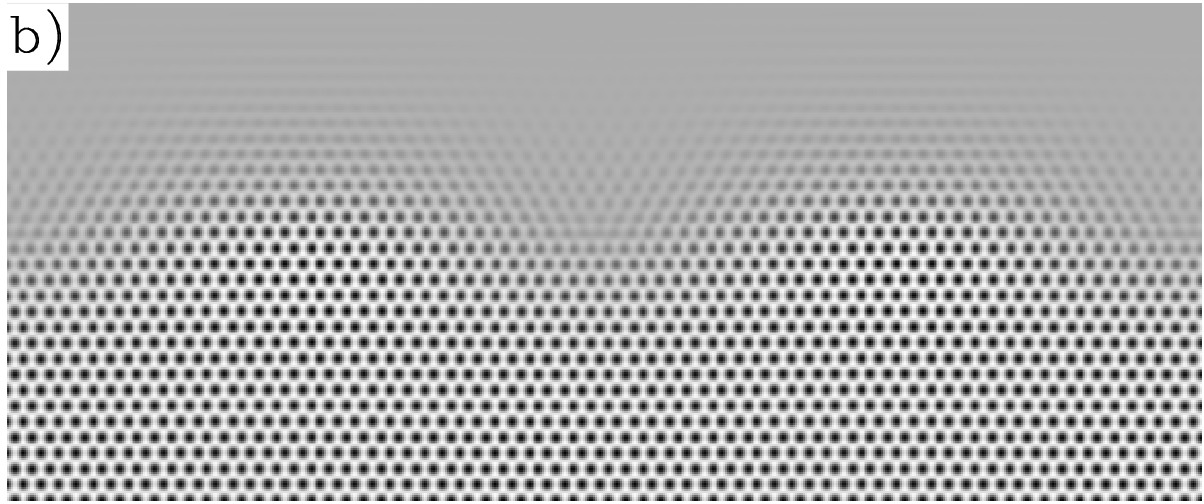}} 
\caption{Reconstruction of full density field $n$ for an 
interface profile showing island formation, with a $3\%$ misfit at
$\epsilon=0.02$.  a) corresponds to density $n$ at $t=125,000$, for  
$B^x=1$ and the maximum instability wave number $q_x=0.0184$, while 
b) corresponds to $n$ at $t=2000$, for $B^x=10$ and $q_x=0.026$.}
\label{fig:configs}
\end{figure}

\begin{figure}
\centerline{
\includegraphics[height=2.3in]{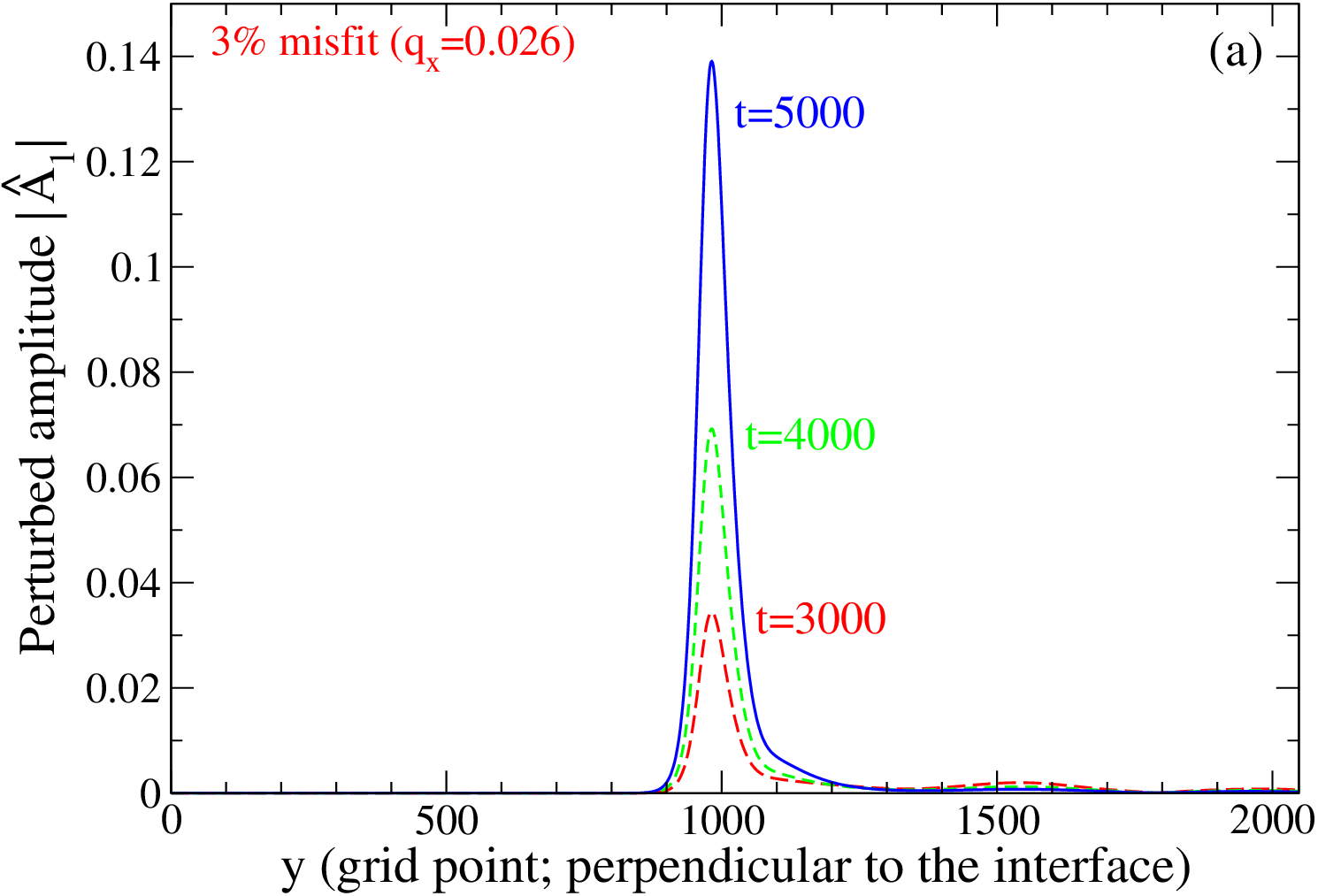}}
\vskip 26pt
\centerline{
\includegraphics[height=2.26in]{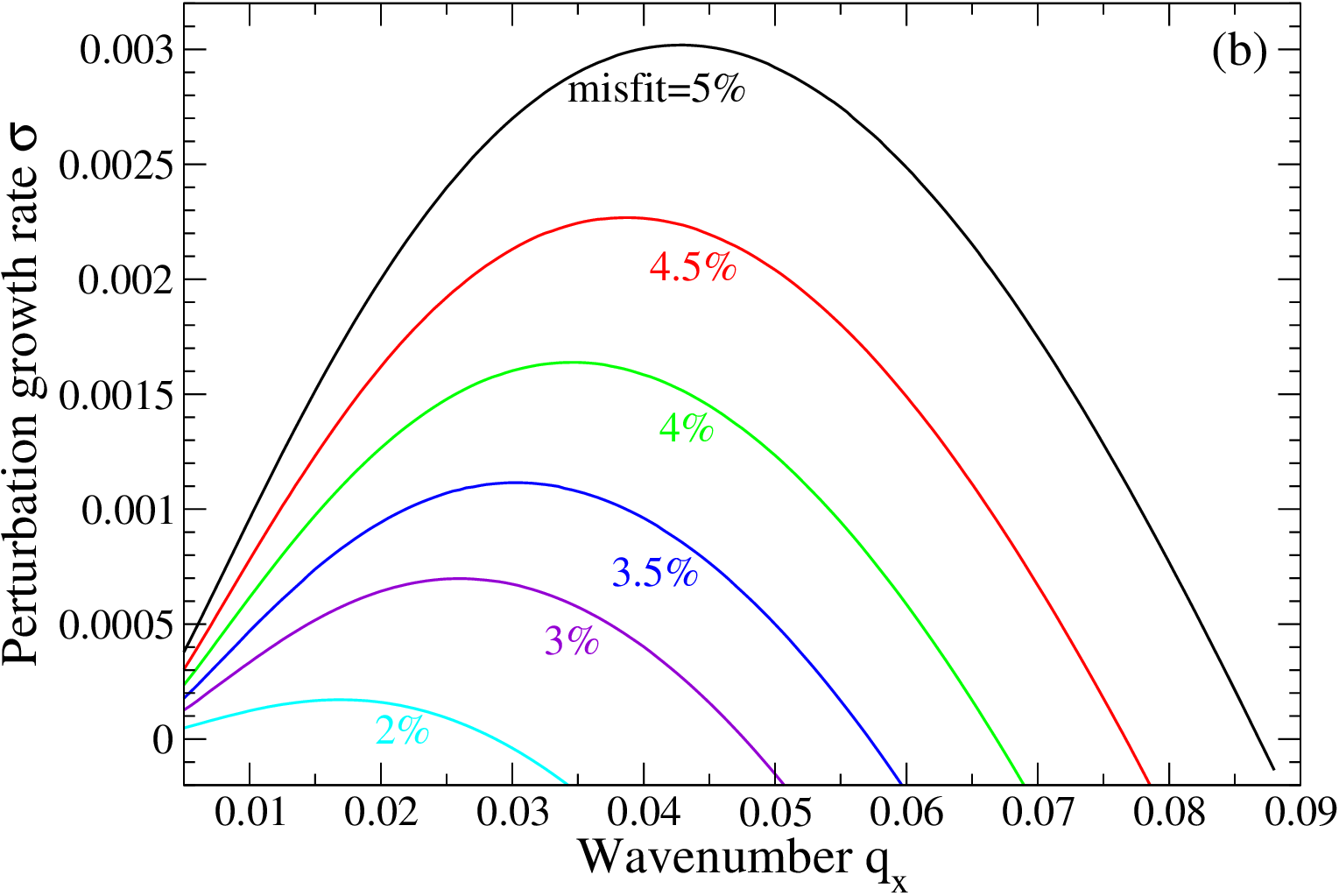} \hfill
\includegraphics[height=2.3in]{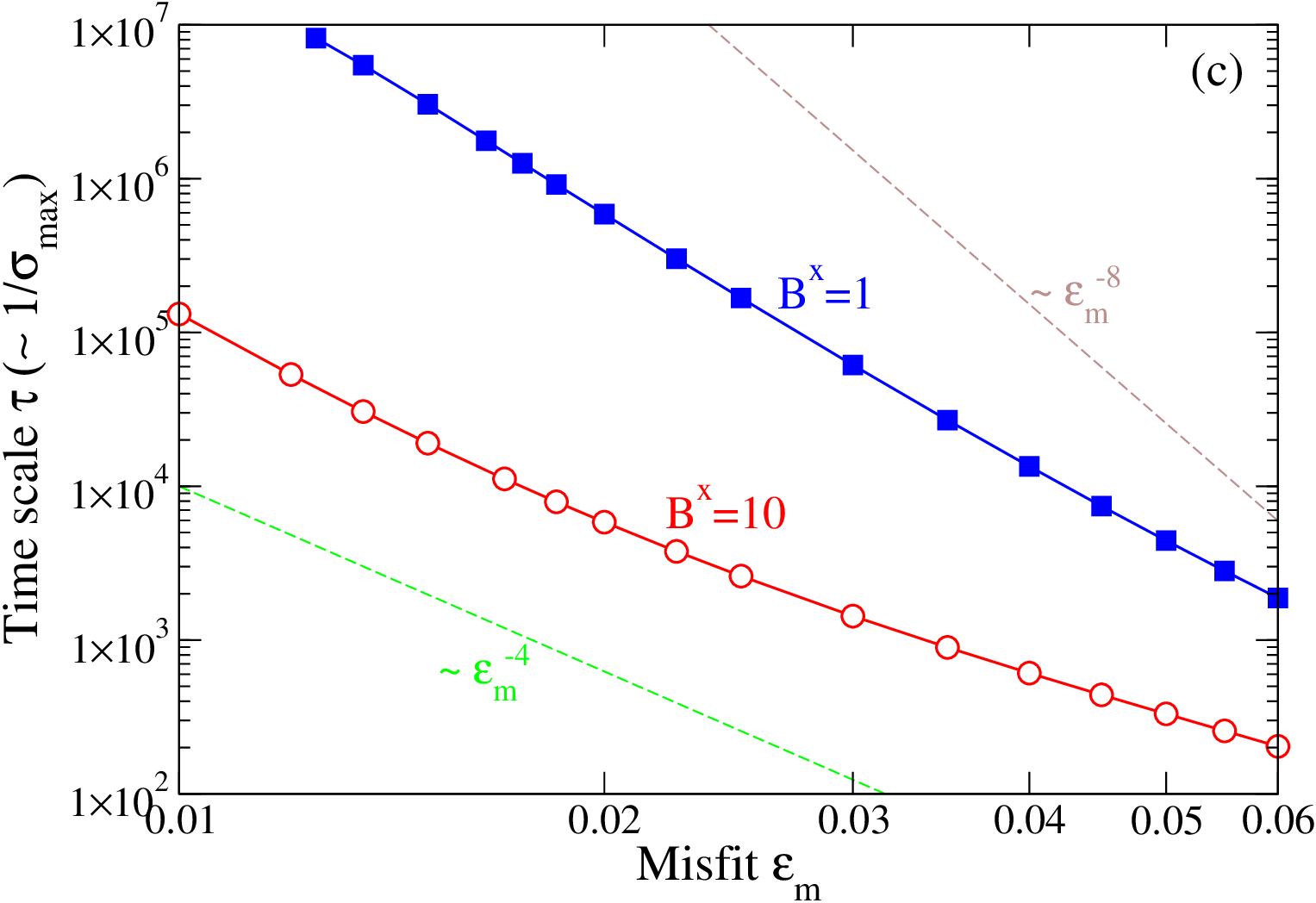}}
\caption{(a) Amplitude perturbations, which grow with time around the
  solid/liquid interface for $\epsilon=0.02$, $B^x=10$, wave number 
  $q_x=0.026$ and $3\%$ misfit.
(b) Perturbation growth rate $\sigma$ as a function of wave number
  $q_x$, for different values of misfit $\varepsilon_m$. Other 
  parameters are the same as (a).
(c) Characteristic time scale $\tau$ ($\sim 1/\sigma_{\rm max}$) for 
  the mounding instability as a function of misfit $\varepsilon_m$, 
  for $B^x=1$ and $10$. Two power laws, $\tau \sim \varepsilon_m^{-8}$
  and $\sim \varepsilon_m^{-4}$, are also shown for comparison.}
\label{fig:perb_sig}
\end{figure}
 
For nonzero misfit, within a certain range of wave number $q_x$
the initial perturbations of $\hat{A}_j$ and $\hat{n}_0$ grow
with time around the liquid-solid interface, while they always 
decay to zero far from the interface region, showing the stability
of both the solid and liquid bulks.  This interface instability 
results in the formation of islands or mounds at the liquid-solid
interface, as shown in Fig. \ref{fig:configs}. This figure was 
obtained by reconstructing full density field $n$ from the amplitudes
with wave number $q_x$ of maximum instability (based on
Eq. (\ref{eq:nexpan})). A typical example of the dynamics of the
amplitudes that gives rise to this instability is given in 
Fig. \ref{fig:perb_sig}a.
We then calculate the perturbation growth rate $\sigma(q_x)$,
noting that $|\hat{A}_j|, |\hat{n}_0| \propto e^{\sigma t}$. This
process is repeated for a range of perturbation wave number $q_x$, and
also for various misfits $\varepsilon_m$. Some results of the dispersion
relation are shown in Fig. \ref{fig:perb_sig}b, for $\epsilon=0.02$
and $B^x=10$. Previous work of continuum elasticity or phase-field
theory has predicted various forms of dispersion relation, including
$\sigma \simeq \alpha_3 q^3 - \alpha_4 q^4$ (for surface-diffusion 
dominated process, \cite{re:srolovitz89,re:spencer91,re:spencer93})
$\sigma \simeq -\alpha_2 q^2 +\alpha_3 q^3 - \alpha_4 q^4$ (if considering 
wetting effects, \cite{re:levine07,re:eisenberg00})
$\sigma \simeq \alpha_1 q - \alpha_2 q^2$ (in the case of 
evaporation-condensation, \cite{re:srolovitz89,re:muller99,re:kassner01})
or $\sigma \simeq \alpha_2 q^2 - \alpha_3 q^3$ (for bulk-diffusion 
dominated case, \cite{re:wu09})
with $q$ the wave number and $\alpha_i$ ($i=1,...,4$) the model-dependent
coefficients that are usually a function of surface tension and elastic
moduli. However, none of these forms fits our dispersion data, which 
instead can be well fitted only by a 4th order polynomial of $q_x$ for
all range of wave numbers, similar to a combination of all the above forms.
This is not unexpected, given that all factors of surface diffusion,
bulk diffusion, wetting effects, and evaporation/condensation are naturally
incorporated in the PFC model and cannot be easily decoupled. This can
be seen through the fact that the PFC modeling of epitaxial growth
involves the coexistence of liquid-solid interface that buckles
and evolves, and thus naturally involves the diffusion processes along
the interface and between liquid region and solid film, and also the
variation of material properties such as surface/interface energy and
elastic relaxation across the interface (i.e., the wetting effects). 
We expect that an important parameter controlling these different
processes would be $\epsilon$, the temperature distance from the melting
point. The $\epsilon$ (or temperature) dependence of properties of
system relaxation has been known for pattern formation systems, and 
is also seen in our PFC studies. Here we focus on high
temperature regime where the amplitude equation representation is most
relevant and effective, and hence choose $\epsilon=0.02$ which is
different from other studies with larger $\epsilon$ and hence lower
growth temperature (e.g., $\epsilon=0.1$ in
Ref. \onlinecite{re:wu09}). For such small $\epsilon$ (high
temperature) surface diffusion process is more 
prominent and coupled with the bulk diffusion process, a phenomenon
that might be weakened or absent in low temperature growth (e.g., 
in Ref. \onlinecite{re:wu09} only bulk diffusion behavior has been
identified in the dispersion relation obtained from the original PFC
equation).

The development of surface perturbations and instability can be 
characterized by an evolution time scale $\tau$, which can be approximated
via the inverse of maximum perturbation growth rate $\sigma_{\rm max}$
and is found to scale as $\varepsilon_m^{-8}$ or $\varepsilon_m^{-4}$ 
in continuum elasticity theory with the assumed mass transport mechanism 
dominated by surface diffusion or evaporation-condensation respectively.
\cite{re:srolovitz89,re:spencer93} However, our calculations yield 
results more complicated than this single power law behavior, as shown in 
Fig. \ref{fig:perb_sig}c, which can also be expected from the coupling of
various mass transport processes in this modeling as discussed above.
Our results show that the time scale $\tau$ decreases with misfit 
strain $\varepsilon_m$, since the $\varepsilon_m$ provides the driving 
force for the morphological instability.   $\tau$ is also 
found to significantly decreases when $B^x$ increases.  For example
at a given misfit, $\tau$ is typically one or two orders of magnitude 
larger for $B^x=1$ compared with $B^x=10$.   This difference is 
most likely due to the significant decrease in surface energy and 
increase in interfacial thickness as $B^x$ is increased, as shown 
in Fig. \ref{fig:gamma} and Fig. \ref{fig:A0n0} respectively.

\begin{figure}
\vskip 0.3in
\centerline{
\includegraphics[height=2.3in]{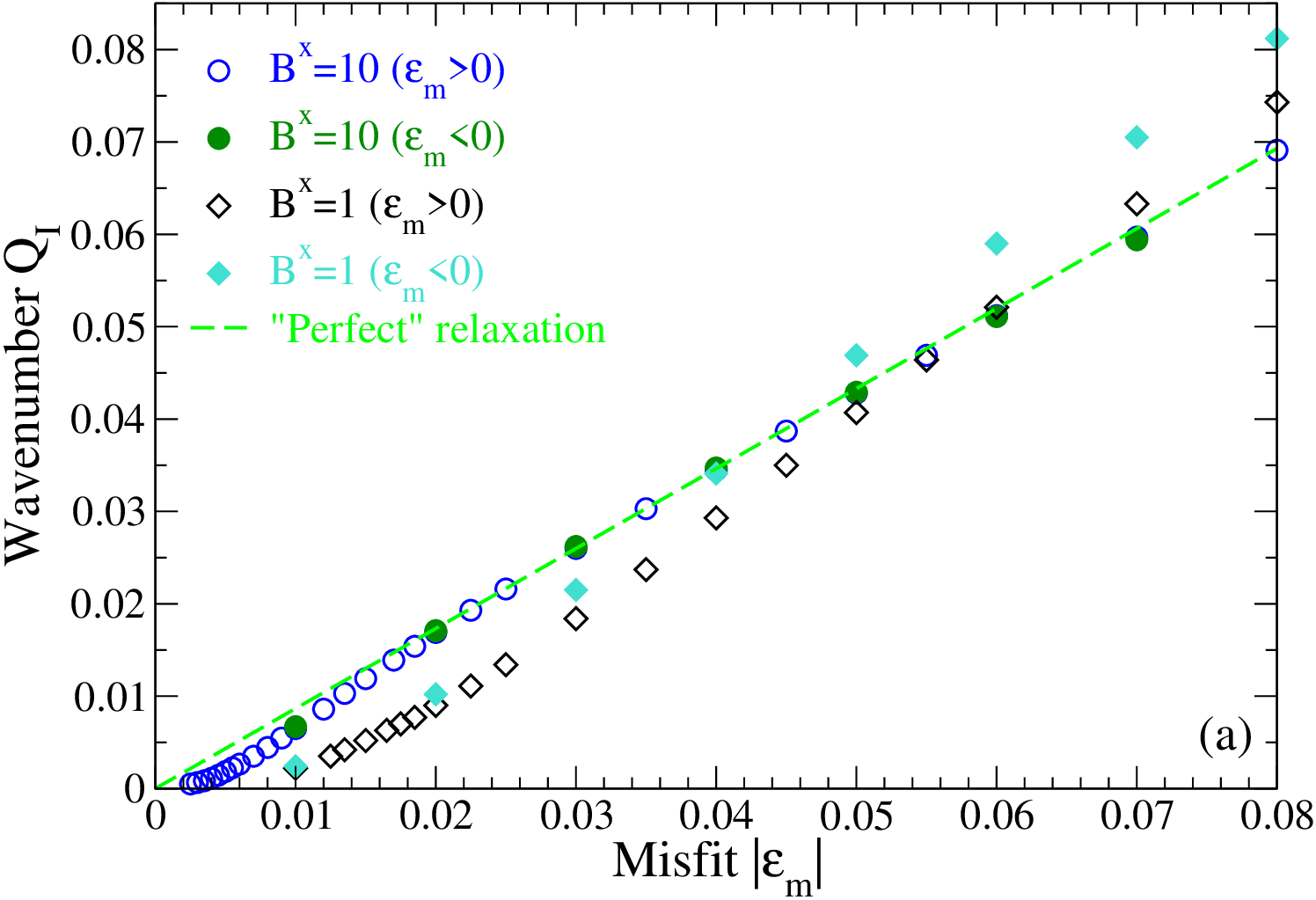} \hfill
\includegraphics[height=2.3in]{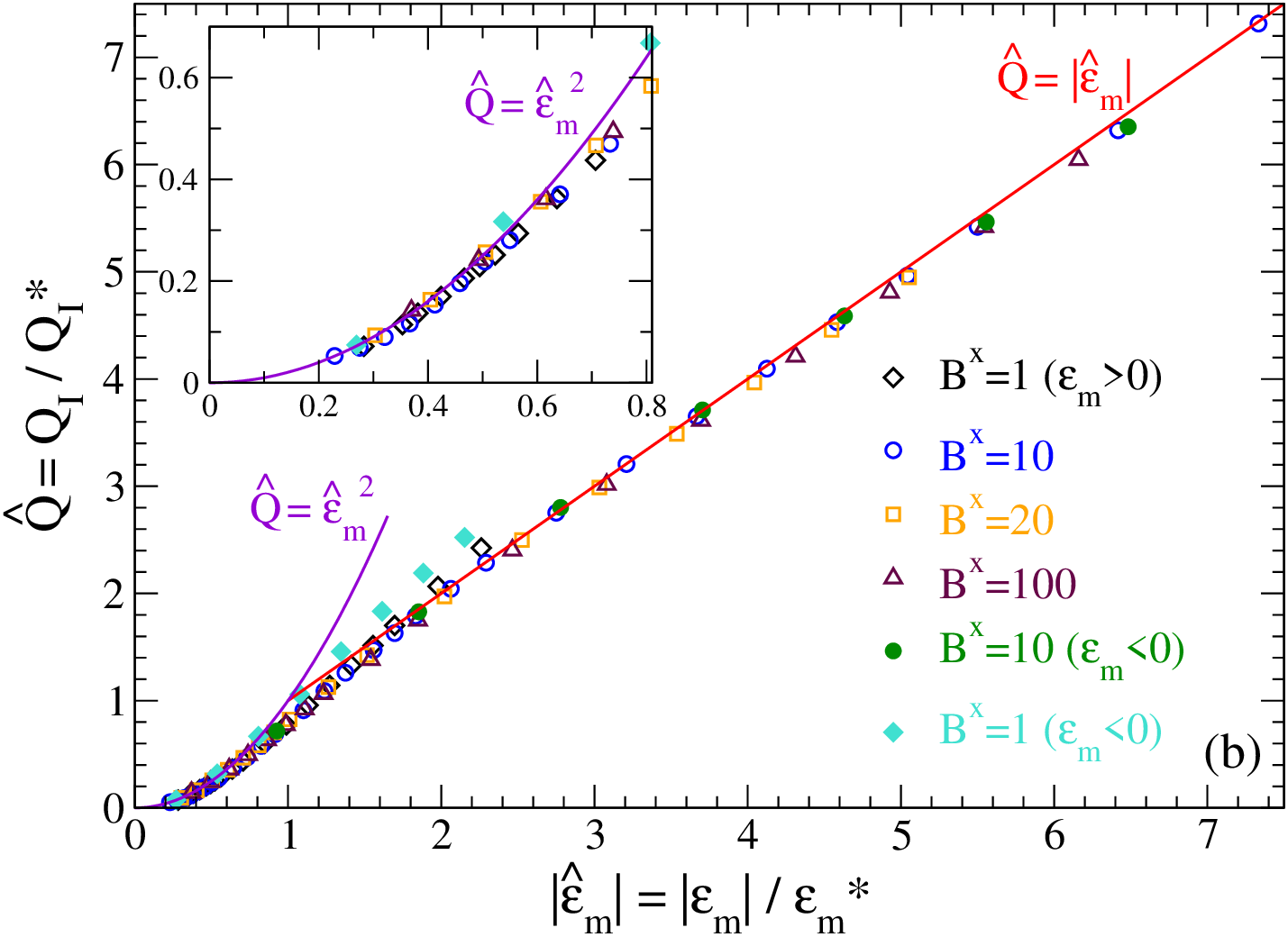}}
\caption{(a) Characteristic wave number $Q_I$ of film surface instability
  as a function of misfit strain magnitude $|\varepsilon_m|$, for
  different values of $B^x=1$ and $10$ and both compressive
  ($\varepsilon_m>0$) and tensile ($\varepsilon_m<0$) films. The limit
  imposed by ``perfect'' relaxation condition is indicated by a dashed line.
  (b) Scaling of island wave number based on a crossover wave number
  $Q^*=3\gamma q_0^2/4E$ and misfit $\varepsilon_m^*=3\gamma q_0/4E$,
  for different values of $B^x$ which is proportional to film elastic
  modulus. The inset highlights the crossover to the continuum result
  of $Q_I \sim \varepsilon_m^2$ at small misfit limit.}
\label{fig:Q_misfit}
\end{figure}

The maximum of the growth rate determines the characteristic wave number
$Q_I$ for the instability, and hence the characteristic wave number
of the island/mound formation on the film surface. 
We plot in Fig. \ref{fig:Q_misfit}a the relation of this instability/island 
wave number $Q_I$ vs. misfit strain $\varepsilon_m$, for different values
of $B^x$ and for both compressive ($\varepsilon_m>0$) and tensile
($\varepsilon_m<0$) films.  For each value of $B^x$ we can identify 
two regions, corresponding to a quadratic behavior 
of $Q_I \sim \varepsilon_m^2$ 
at small misfits (see also the inset of Fig. \ref{fig:Q_misfit}b) and a 
linear dependence of $Q_I$ on $\varepsilon_m$ for large enough strains.
Such quadratic scaling in the small misfit limit is consistent with
the well-known results of continuum theory including all different 
assumptions of dominant mechanisms such as surface diffusion, 
evaporation-condensation and wetting effects.
\cite{re:srolovitz89,re:spencer91,re:spencer93,re:spencer05,re:levine07,re:eisenberg00}
However, this $\varepsilon_m^2$ scaling result differs from the
experimental findings in SiGe/Si(001) growth, \cite{re:sutter00,re:tromp00}
which indicate the linear behavior $Q_I \sim \varepsilon_m$ for the
stress-driven surface instability and coherent epitaxial islands.
Although this observation of a linear relationship is qualitatively
similar to what we obtain above for large enough misfits, 
it should be cautioned that the experimental systems involve more
complicated factors related to the SiGe alloying nature that is not considered
here, particularly the atomic mobility difference between the two film 
components which was verified by recent first principle calculations 
\cite{re:huangl06} and was believed to play a key role on island size
scaling. \cite{re:tersoff00,re:spencer01}

For the single-component films studied here the crossover from 
the quadratic scaling at the continuum weak-strain limit to 
linear behavior at high strains is most likely due to the 
discrete nature of the crystalline lattice that is implicitly 
included in the amplitude formulation.
It is known (and verified in direct simulations of PFC
Eq. (\ref{eq:pfc}) \cite{re:elder04,re:elder07,re:huang08})
that at late times the instability to form islands or mounds leads 
to the nucleation of dislocations around the edges of islands 
or in the valleys between the mounds.  These dislocations nucleate 
to relieve strain in the film and appear at earlier times for larger 
misfit strains.  Here we define a length scale, $\lambda_R$, 
for ``perfect'' relaxation such that if the dislocations nucleate 
at this distance apart, strain in the film will be completely relieved 
(aside from the strain induced by the dislocations themselves).
We can then make the assumption that if the continuum prediction for 
most unstable wavelength is smaller than $\lambda_R$, continuum 
theory will break down.  To evaluate $\lambda_R$ consider 
a 1+1 dimensional film; assume $L_x$ being 
the lateral length of film surface and by definition we have 
$L_x = N a = M a_0$, where $N$ is the number of atoms in strained lattice,
$M$ is the atom number for unstrained state after dislocations nucleate,
and $a$ and $a_0$ are the corresponding lattice constants already defined 
in Eq. (\ref{eq:misfit}). Thus from Eq. (\ref{eq:misfit}) for the definition 
of misfit, we obtain $\varepsilon_m=(N-M)/M$, leading to the average 
distance between dislocations $\lambda_R = L/|N-M| = L/(M|\varepsilon_m|)
= a_0 /|\varepsilon_m|$, with the associated wave number $Q_R=q_{x0} 
|\varepsilon_m|$ (plotted as a dashed line in Fig. \ref{fig:Q_misfit}a).
Assuming that on average at least one dislocation will appear at
each island edge/valley, this wave number $Q_R$ will then be the upper
limit imposed by the discrete nature of the lattice, as it would be
unphysical for islands with size smaller than $\lambda_R$ to appear
which would instead cause the ``overrelaxation" of the film lattice.
Our results of island wave number $Q_I$ for different values of
$B^x$ ($=10, 20, 100$) all converge to this limit at large misfit strains
(except for $B^x=1$ which will be discussed below).

This ``perfect" relaxation condition is expected to be met at large enough
misfits, but not at small strains where dislocations appear at far late
stage after islands form, leading to the crossover phenomenon between two
scaling regimes given in Fig. \ref{fig:Q_misfit}. This crossover occurs
when $Q_I ({\rm of~ small~ misfit~ limit }) = Q_R$. As stated above, at
small $\varepsilon_m$ we can recover the result of continuum theory which
predicts $Q_I \propto (E/\gamma) \varepsilon_m^2$ (with $E$ the Young's
modulus). \cite{re:srolovitz89,re:spencer91,re:spencer93} 
In our calculations based on the PFC model and amplitude equations, 
we evaluate $E$ from a one-mode approximation, \cite{re:elder04,re:elder07}
$E=B^x A_{\rm min}^2/2$, where $A_{\rm min}=4 (g-3n_0 +
\sqrt{g^2+24n_0 g-36 n_0^2 + 15\varepsilon_m})/15$. Using the results of 
$\gamma$ given in Sec. \ref{sec:base}, we can fit the small misfit
data well into a form $Q_I=4E \varepsilon_m^2 / 3\gamma$ (for all values of 
$B^x$; see the inset of Fig. \ref{fig:Q_misfit}b). Therefore, the misfit
($\varepsilon_m^*$) and island wave number ($Q_I^*$) at the crossover
can be determined via $Q_I^*=4E {\varepsilon_m^*}^2 / 3\gamma
=Q_R=q_{x0} \varepsilon_m^*$, resulting in $\varepsilon_m^*=3\gamma q_{x0}/4E$
and $Q_I^* = 3\gamma q_{x0}^2/4E$. Defining rescaled quantities 
$\hat{Q} = Q_I / Q_I^*$ and $\hat{\varepsilon}_m = \varepsilon_m / 
\varepsilon_m^*$, we can then scale all the data from
different conditions (e.g., films of different elastic constants, for
$B^x>1$) onto a single universal scaling curve accommodating all range
of misfit strains, for both compressive and tensile films (see 
Fig. \ref{fig:Q_misfit}b). The crossover misfit strain 
$\varepsilon_m^*$ can be very small ($<2\%$, depending on e.g., film
elastic properties), showing the breakdown of continuum approach even
at relatively large scales.

\begin{figure}
\centerline{
\includegraphics[height=2.5in]{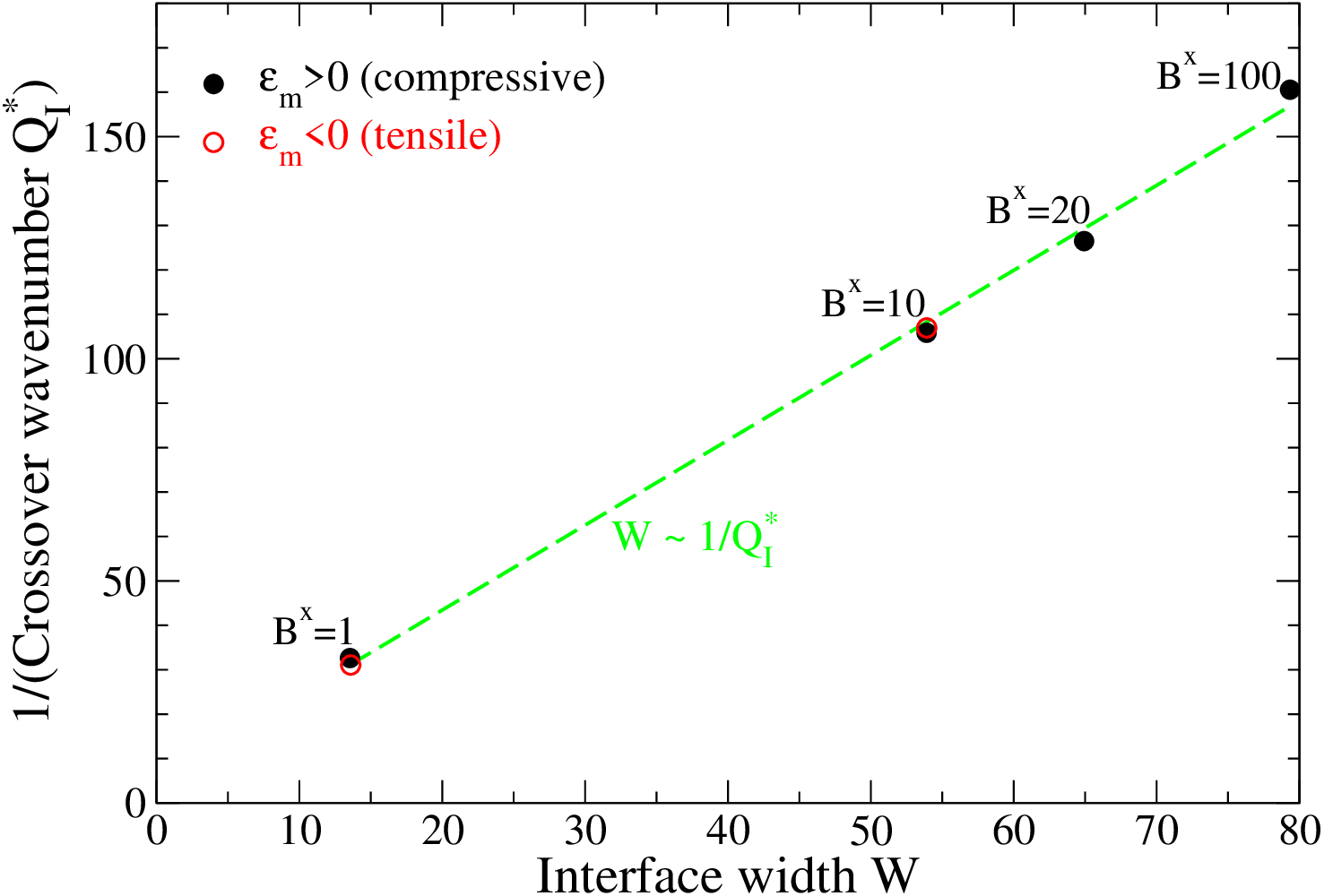}}
\vskip 3pt
\caption{The linear relation between the liquid-solid interface width $W$
and the inverse of crossover wave number $1/Q_I^*$, for $\epsilon=0.02$ and
both compressive (filled symbols) and tensile (open symbols) films. Values
of $B^x$ are also indicated on the plot.}
\label{fig:width}
\end{figure}

Note that although this linear behavior due to ``perfect" lattice
relaxation and the scaling crossover have been observed in our
previous work, \cite{re:huang08} it was limited to compressive
strained films and not-too-large misfits. However, the more generalized
study given here shows a small deviation from the limit of ``perfect"
relaxation for small value of $B^x$, as indicated in Fig. \ref{fig:Q_misfit}a
with island wave numbers of $B^x=1$ lying above such upper limit (the 
dashed line) when the magnitude of mismatch $|\varepsilon_m|$ exceeds $5\%$ 
(for tensile films) or $6\%$ (compressive). Similar deviation can be seen
in the corresponding scaling plot of Fig. \ref{fig:Q_misfit}b. Nevertheless,
at large misfits the linear scaling behavior is still maintained, which is
qualitatively different from the quadratic scaling at the small strain limit.
Based on the discussions given above for ``perfect" relaxation condition,
it is expected that $Q_I > Q_R$ occurs only when some of the island edges
would be dislocation-free even at late evolution times. The condition for
this scenario is not clear; but our results suggest that this may occur
when the liquid-solid interface (or film surface) is sharp enough.
As given in Fig. \ref{fig:width}, the interface width $W$ decreases with
the value of $B^x$, and is particularly small at $B^x=1$ (with $W \sim 
13.5\Delta y$ for both tensile and compressive films, less than 2 lattice 
spacing) as compared to others. It could then be expected that details of 
film morphological evolution, including instability and island formation, 
would be different for such sharp interface, as somewhat indicated in
Fig. \ref{fig:Q_misfit}. Further studies are needed to clarify this
special scenario of strained film evolution.

Fig. \ref{fig:width} also yields the effect of finite interface width 
$W$ on the island size (or wave number) scaling. We find $1/Q_I^* \sim W$,
i.e., a linear relation between crossover instability wavelength
($=2\pi /Q_I^*$) and the interface thickness. This is consistent with
most recent results of direct PFC simulations
\cite{re:wu09} which indicate that the discrepancy or crossover
between the classical elasticity result of quadratic scaling of $Q_I$ 
and the linear behavior identified in the PFC modeling could be attributed
to the finite thickness of the interface, a fact that is neglected in
the classical continuum theory. As seen in Fig. \ref{fig:width}, when
$W \rightarrow 0$ (i.e., the assumption adopted in continuum elasticity
theory), we have $Q_I^* \rightarrow \infty$ and hence recover the
continuum theory prediction of $Q_I \sim \varepsilon_m^2$ for the whole
range of misfit strain, as expected. Corresponding to real
experimental systems, Fig. \ref{fig:width} predicts that at constant
growth temperature (same $\epsilon$ value), the liquid-solid
interface thickness varies with film elastic modulus (or the value of
$B^x$), and for different film materials the crossover island size
separating two island scaling regimes increases linearly with the
interface thickness.

Another important feature of our results is the asymmetry between tensile 
and compressive films which, however, becomes distinct only at small enough 
$B^x$ and large enough misfits (see Fig. \ref{fig:Q_misfit} for the data of 
$B^x=1$). Given the important role played by the surface energy 
$\gamma$ on film stability and evolution, we expect this asymmetric
phenomenon of island wave number to be closely related to the property
of $\gamma$ shown in Fig. \ref{fig:gamma}. The intrinsic surface stress 
$\sigma_{xx}^0$ determined for $B^x=1$ is an order of magnitude larger 
than that for $B^x=10$, leading to much larger value of surface energy
difference between tensile and compressive strains; also such difference
increases with the magnitude of misfit strain. The corresponding behavior
of surface instability and island formation would then follow the similar 
trend, as observed in Fig. \ref{fig:Q_misfit}.

\section{Free Energy Analysis and Mode Coupling}

To further elucidate the properties of the strained surface, it 
is interesting to analyze the effective free 
energy ${\cal F}$ (given in Eq. (\ref{eq:F})).  Consider the 
net change of ${\cal F}$ relative to that of a planar
interface, i.e., 
\begin{equation}
\Delta {\cal F} = {\cal F} - {\cal F}^0,
\end{equation}
where ${\cal F}^0$ is the free energy of the planar interface given in
Eq. (\ref{eq:F0}). $\Delta {\cal F} <0$ indicates film surface
instability against the initial perturbation, while $\Delta {\cal F} >0$
refers to the energy penalty of any perturbations and hence
corresponds to stability of planar film surface.

Based on the Fourier expansion (\ref{eq:A_expan}) and
(\ref{eq:n_expan}), $\Delta {\cal F}$ can be expanded up to second
order in the perturbed quantities $\hat{A}_j$ and $\hat{n}_0$, i.e.,
\begin{equation}
\Delta {\cal F} = \Delta {\cal F}^{(1)} + \Delta {\cal F}^{(2)}.
\label{eq:dF12}
\end{equation}
Detailed expression of the first order term $\Delta {\cal F}^{(1)}$ is
given in the Appendix (see Eq. (\ref{eq:dF1})).
We find numerically $\Delta {\cal F}^{(1)} \sim 0$, 
and hence the net energy change $\Delta {\cal F}$ is determined by the
second order quantity
\begin{equation}
\Delta {\cal F}^{(2)} = \Delta {\cal F}_{+} + \Delta {\cal F}_{-},
\end{equation}
where
\begin{eqnarray}
&\Delta {\cal F}_{-} = L_x \int dy \sum\limits_{q_x} \left \{
  (6n_0^0 - 2g) \sum\limits_{j=1}^{3} \left [ A_j^0 \hat{A}_j^*(-q_x)
  + {A_j^0}^* \hat{A}_j(q_x) \right ] \hat{n}_0^*(q_x)
  \right. & \nonumber\\
& \left. + (6n_0^0 - 2g) \left [ A_3^0 \hat{A}_1(q_x) \hat{A}_2(-q_x) +
  A_2^0 \hat{A}_1(q_x) \hat{A}_3(-q_x) + A_1^0 \hat{A}_2(q_x)
  \hat{A}_3(-q_x) + {\rm c.c.} \right ] \right \}, & \nonumber\\
&& \label{eq:dF_}
\end{eqnarray}
with $\hat{A}_j(q_x) = \hat{A}_j(q_x,y,t)$ and $\hat{n}_0(q_x) =
\hat{n}_0(q_x,y,t)$, and the contribution $\Delta {\cal F}_{+}$ is
shown in Eq. (\ref{eq:dF+}) of the Appendix.
 
\begin{figure} 
\centerline{ 
\includegraphics[height=2.8in]{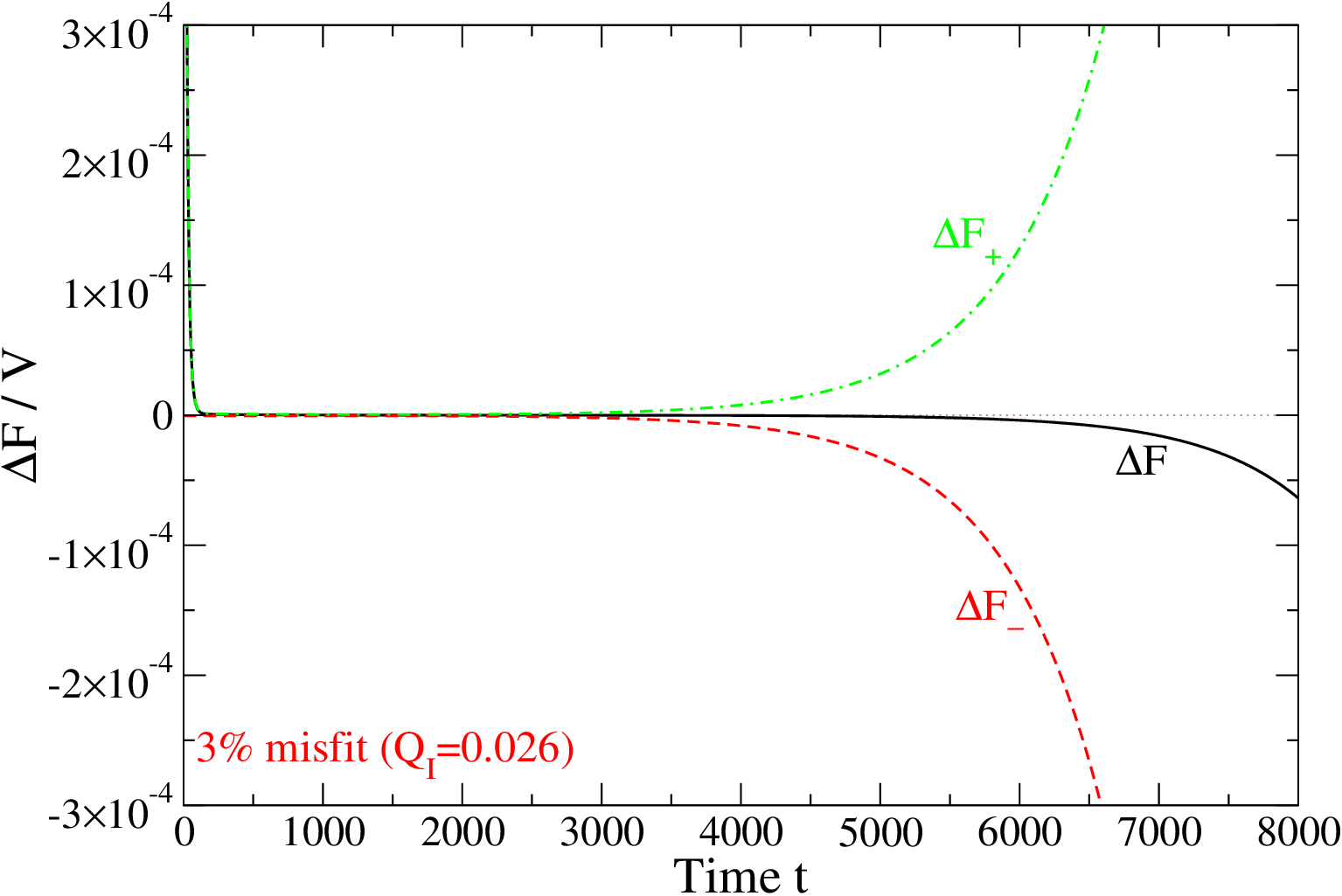}}
\vskip 3pt
\caption{Time evolution of effective free energy density change 
    $\Delta {\cal F}$ (per unit volume $V=L_x L_y$) of the perturbed state, 
    with misfit $\varepsilon_m=3\%$, wave number $Q_I=0.026$, $\epsilon=0.02$, 
    and $B^x=10$. Also included are the positive contribution
    $\Delta {\cal F}_+$ and the negative contribution $\Delta {\cal
    F}_-$.}
\label{fig:perb_dF} 
\end{figure}

Given the numerical solution for the perturbed amplitudes (see
Eqs. (\ref{eq:ampl_hatA1})--(\ref{eq:ampl_hatn0})) as described
in Sec. \ref{sec:perb}, $\Delta {\cal F}$ ($\simeq \Delta
{\cal F}^{(2)}$) can be approximated via the most unstable characteristic 
wave number by substituting the numerical solutions for amplitudes 
at $q_x=\pm Q_I$.
We find that all terms in
Eq. (\ref{eq:dF+}) are positive, i.e., $\Delta {\cal F}_{+} > 0$;
both two terms in Eq. (\ref{eq:dF_}) yield negative contribution
(noting that usually $6n_0^0 - 2g <0$ for liquid-solid coexistence), 
so that $\Delta {\cal F}_{-} < 0$, and the magnitude of the last term 
is much larger than the 1st one.  As shown in Fig. \ref{fig:perb_dF},
at large enough time $\Delta {\cal F}_{-}$ dominates over the
stabilizing terms in $\Delta {\cal F}_{+}$, leading to negative net 
free energy change $\Delta {\cal F}$ and thus the film instability. 
Note that the last term in Eq. (\ref{eq:dF_}), which dominates 
$\Delta {\cal F}_{-}$, arises from the 2nd-order expansion of 
$(A_1 A_2 A_3 + A_1^* A_2^* A_3^*)$ in the effective free energy
formula (\ref{eq:F}). It represents the coupling of different modes 
of complex amplitudes, and our numerical results show that it 
contributes to the integral of $\Delta {\cal F}_-$ only in the interface 
or film surface region (as the perturbed amplitudes decay fast in the bulks).
We can then argue that it is the mode coupling of complex amplitudes at
the liquid-solid interface that is mainly responsible for the
morphological instability of the strained film.
Note that the amplitudes of structural profile $A_j$ are complex, and
thus their evolution involves an important process of phase
perturbation (or phase winding). Physically this phase behavior
corresponds to the elastic relaxation of the lattice structure, and
thus the mode coupling property identified above indicates that the
coupling of elastic relaxation for different lattice modes (or
wave vectors) around the film surface would be one of the major factors
underlying the film instability and mounding behavior. Such phase
behavior is related to details of crystalline structure, as captured
by the PFC model and the amplitude equation formalism, but not by the
continuum theory. Furthermore, the competition between 
$\Delta {\cal F}_+$ ($>0$) and $\Delta {\cal F}_-$ ($<0$) shown in
Fig. \ref{fig:perb_dF} is consistent with previous analysis of
continuum elasticity theory showing the competition between film
stabilization effects (such as surface energy) and destabilizing
factors (mainly elastic effects). \cite{re:srolovitz89,re:spencer91,%
re:spencer93,re:guyer95,re:leonard98,re:spencer01,re:huang02a,re:huang02b}
Note also that the above mechanism identified should
be already incorporated in the original PFC equation (\ref{eq:pfc})
and the associated PFC free energy (\ref{eq:pfc_F}), while the
analysis given here based on the amplitude formulation has the
advantage of being able to single out individual contributions from
different lattice modes.

\section{Conclusions}

We have investigated the detailed properties of a strained film surface,
its morphological instability, and the associated island wave number 
scaling through a systematic analysis of the amplitude equation
formalism based on the phase field crystal model.
We identify the amplitude and average density profiles of
liquid-film coexisting interface, the interface width, miscibility gap,
and surface energy (including intrinsic surface stress and excess elastic
modulus), for various misfit strains (both magnitude and sign) and film 
elastic constants (or values of $B^x$). The morphological or mounding
instability of the strained film is systematically examined, showing
results absent in all previous continuum elasticity and phase-field 
approaches and atomistic modeling. In particular,
we obtain a crossover phenomenon of instability or island wave number 
scaling, from the well-known continuum, ATG result of $Q_I \sim 
\varepsilon_m^2$ to a linear behavior $Q_I \sim \varepsilon_m$
at large enough strains which is identified by an upper limit imposed
by the condition of ``perfect" lattice relaxation. Most data (of different
parameter ranges) can be scaled onto a universal scaling relation for
the whole range of misfit strain, with some small deviations for 
very narrow liquid-solid interfaces in the large strain limit.
The asymmetry of film properties between tensile and compressive 
strains is also observed. Note that although either linear or 
quadratic scaling has been reported in
experiments (such as SiGe/Si(001)) and model simulations (e.g.,
kinetic MC) or continuum theory (e.g., ATG instability), the universal
scaling relation with crossover of the two regions has not been found
before. We expect our prediction here to be examined by experiments of
single-component film epitaxy or atomistic simulations with large
enough length and time scales.

Our study highlights an important feature of the amplitude formulation
for strained film epitaxy, in that it can simultaneously 
reproduce continuum results (e.g., the ATG instability) and reveal significant 
corrections due to the microscopic nature of the crystalline structure.
Our approach adopts a mesoscopic-level description of the system,
via the amplitudes or envelopes of the slowly varying surface profile for
which the well-developed continuum, mesoscopic theory can be applied.
On the other hand, the crystalline nature of the strained film is preserved
particularly via phase perturbations of the complex amplitudes that are
prominent around the film surface. The latter has been emphasized through
revealing the breakdown of traditional continuum approaches even at 
relatively small misfit stress and the associated crossover effect of 
island size scaling, and also through examining the origin of film
instability that is accompanied by mode coupling of complex
amplitudes in the liquid-solid interface region. Our results thus emphasize
the importance of multiple scale modeling of complex material systems such
as the strained film epitaxy process studied above. Note that although
in this paper we focus on 2D hexagonal/triangular crystal structure,
we expect the approach and analysis technique developed here to be
directly extended for other crystalline symmetries and other surface
directions, such as the epitaxial growth and island formation in 3D
bcc or fcc films for which we have developed the corresponding
amplitude expansion formulation very recently. \cite{re:elder10}

\begin{acknowledgments}

We are indebted to Kuo-An Wu and Peter Voorhees for helpful discussions.
This work was supported by the National Science Foundation under
Grant No. CAREER DMR-0845264 (Z.-F.H.) and DMR-0906676 (K.R.E.).

\end{acknowledgments}

\appendix
\section{Free energy expansion}
\label{append}

In this appendix the detailed expansion forms of free energy
difference $\Delta {\cal F}$ are presented. For the first order term
$\Delta {\cal F}^{(1)}$ shown in Eq. (\ref{eq:dF12}), we have
\begin{eqnarray}
&\Delta {\cal F}^{(1)}& = L_x \int dy \left \{ \sum_{j=1}^{3}
    \left ( -\epsilon + 3{n_0^0}^2 - 2g n_0^0 + 3 |A_j^0|^2 \right )
    \left ( {A_j^0}^* \hat{A}_j(0) + {\rm c.c.} \right ) 
    \right. \nonumber\\
&&  + (6n_0^0 - 2g) \sum_{j=1}^{3} |A_j^0|^2 \hat{n}_0(0) \nonumber\\
&& + \left [ \left ( \partial_y^2 + i(q_0+\delta_y) \partial_y -\sqrt{3} q_0
    \delta_x - \delta_x^2 - q_0 \delta_y /2 - \delta_y^2/4 \right )
    {A_1^0}^* \right ] \nonumber\\
&& \times \left [ \left ( \partial_y^2 - i(q_0+\delta_y) \partial_y
    -\sqrt{3} q_0 \delta_x - \delta_x^2 - q_0 \delta_y /2 -
    \delta_y^2/4 \right ) \hat{A}_1(0) \right ] + {\rm c.c.}
    \nonumber\\
&& + \left [ \left ( \partial_y^2 - 2i(q_0+\delta_y) \partial_y - 2q_0
    \delta_y - \delta_y^2 \right ) {A_2^0}^* \right ] \nonumber\\
&& \times \left [ \left ( \partial_y^2 + 2i(q_0+\delta_y)
    \partial_y - 2q_0 \delta_y - \delta_y^2
    \right ) \hat{A}_2(0) \right ] + {\rm c.c.}\nonumber\\
&& + \left [ \left ( \partial_y^2 + i(q_0+\delta_y) \partial_y -\sqrt{3} q_0
    \delta_x - \delta_x^2 - q_0 \delta_y /2 - \delta_y^2/4 \right )
    {A_3^0}^* \right ] \nonumber\\
&& \times \left [ \left ( \partial_y^2 - i(q_0+\delta_y) \partial_y
    -\sqrt{3} q_0 \delta_x - \delta_x^2 - q_0 \delta_y /2 -
    \delta_y^2/4 \right ) \hat{A}_3(0) \right ] + {\rm c.c.}
    \nonumber\\
&& + (6n_0^0 - 2g) \left [ A_2^0 A_3^0 \hat{A}_1(0) + A_1^0 A_3^0
    \hat{A}_2(0) + A_1^0 A_2^0 \hat{A}_3(0) + {\rm c.c.} \right ]
    \nonumber\\
&& + 6 \left (A_1^0 A_2^0 A_3^0 + {A_1^0}^* {A_2^0}^* {A_3^0}^* \right
    ) \hat{n}_0(0) + 6 \left ( |A_2^0|^2 + |A_3^0|^2 \right )
    \left ( {A_1^0}^* \hat{A}_1(0) + {\rm c.c.} \right ) \nonumber\\
&& + 6 \left ( |A_1^0|^2 + |A_3^0|^2 \right ) \left ( {A_2^0}^*
    \hat{A}_2(0) + {\rm c.c.} \right ) + 6 \left ( |A_1^0|^2 + |A_2^0|^2 \right )
    \left ( {A_3^0}^* \hat{A}_3(0) + {\rm c.c.} \right ) \nonumber\\
&& \left. + \left ( -\epsilon + {n_0^0}^2 - g n_0^0 \right ) n_0^0
    \hat{n}_0(0) + \left [ \left ( \partial_y^2 + q_0^2 \right ) n_0^0
    \right ] \left [ \left ( \partial_y^2 + q_0^2 \right )
    \hat{n}_0(0) \right ] \right \},
\label{eq:dF1}
\end{eqnarray}
with $\hat{A}_j(0) = \hat{A}_j(q_x=0,y,t)$ and $\hat{n}_0(0) =
\hat{n}_0(q_x=0,y,t)$. For the second order terms, the contribution 
$\Delta {\cal F}_{+}$ is given by
\begin{eqnarray}
&\Delta {\cal F}_{+}& = L_x \int dy \sum\limits_{q_x} \left \{
  \sum_{j=1}^{3} \left ( -\epsilon + 3{n_0^0}^2 - 2g n_0^0 + 3
  |A_j^0|^2 \right ) |\hat{A}_j(q_x)|^2 \right. \nonumber\\
&& + \frac{3}{2} \sum_{j=1}^{3} \left | {A_j^0}^* \hat{A}_j(q_x) +
  A_j^0 \hat{A}_j^*(-q_x) \right |^2 \nonumber\\
&& + \left | \left [ \partial_y^2 - i(q_0+\delta_y) \partial_y - q_x^2
  + (\sqrt{3}q_0 + 2\delta_x ) q_x \right. \right. \nonumber\\ 
&& \left. \left. -\sqrt{3} q_0
    \delta_x - \delta_x^2 - q_0 \delta_y /2 - \delta_y^2/4 \right ]
  \hat{A}_1(q_x) \right |^2 \nonumber\\
&& + \left | \left [ \partial_y^2 + 2i(q_0+\delta_y)
    \partial_y - q_x^2 - 2q_0 \delta_y - \delta_y^2 \right ]
  \hat{A}_2(q_x) \right |^2 \nonumber\\
&& + \left | \left [ \partial_y^2 - i(q_0+\delta_y) \partial_y - q_x^2
  - (\sqrt{3}q_0 + 2\delta_x ) q_x \right. \right. \nonumber\\ 
&& \left. \left. -\sqrt{3} q_0
    \delta_x - \delta_x^2 - q_0 \delta_y /2 - \delta_y^2/4 \right ]
  \hat{A}_3(q_x) \right |^2 \nonumber\\
&& + 6 \left [ \left ( |A_2^0|^2 + |A_3^0|^2 \right )
  |\hat{A}_1(q_x)|^2 + \left ( |A_1^0|^2 + |A_3^0|^2 \right )
  |\hat{A}_2(q_x)|^2 \right. \nonumber\\
&& \left. + \left ( |A_1^0|^2 + |A_2^0|^2 \right )
  |\hat{A}_3(q_x)|^2 \right ] \nonumber\\
&& + 6 \left [ \left ( A_2^0 A_3^0 \hat{A}_1(q_x) + A_1^0 A_3^0
  \hat{A}_2(q_x) + A_1^0 A_2^0 \hat{A}_3(q_x) \right )
  \hat{n}_0^*(q_x) + {\rm c.c.} \right ] \nonumber\\
&& + 6 \left [ \left ( {A_1^0}^* \hat{A}_1(q_x) + A_1^0
  \hat{A}_1^*(-q_x) \right ) \left ( {A_2^0}^* \hat{A}_2(-q_x) + A_2^0
  \hat{A}_2^*(q_x) \right ) \right. \nonumber\\
&& + \left ( {A_1^0}^* \hat{A}_1(q_x) + A_1^0
  \hat{A}_1^*(-q_x) \right ) \left ( {A_3^0}^* \hat{A}_3(-q_x) + A_3^0
  \hat{A}_3^*(q_x) \right ) \nonumber\\
&& \left. + \left ( {A_2^0}^* \hat{A}_2(q_x) + A_2^0
  \hat{A}_2^*(-q_x) \right ) \left ( {A_3^0}^* \hat{A}_3(-q_x) + A_3^0
  \hat{A}_3^*(q_x) \right ) \right ] \label{eq:dF+} \\
&& \left. + \frac{1}{2} \left [ -\epsilon + 3{n_0^0}^2 - 2g n_0^0 + 6
  \sum_{j=1}^{3} |A_j^0|^2 \right ] |\hat{n}_0(q_x)|^2 + \frac{1}{2}
  \left | \left ( \partial_y^2 -q_x^2 + q_0^2 \right ) \hat{n}_0(q_x)
  \right |^2 \right \}. \nonumber
\end{eqnarray}


\end{document}